# Annihilation of Relative Equilibria in the Gravitational Field of Irregular-shaped Minor Celestial Bodies

Yu Jiang[1, 2], Hexi Baoyin[2]

1. State Key Laboratory of Astronautic Dynamics, Xi'an Satellite Control Center, Xi'an 710043, China

2. School of Aerospace Engineering, Tsinghua University, Beijing 100084, China

Y. Jiang (✉) e-mail: jiangyu_xian_china@163.com (corresponding author)

**Abstract**. The rotational speeds of irregular-shaped minor celestial bodies can be changed by the YORP effect. This variation in speed can make the numbers, positions, stabilities, and topological cases of the minor body's relative equilibrium points vary. The numbers of relative equilibrium points can be reduced through the collision and annihilation of relative equilibrium points, or increase through the creation and separation of relative equilibrium points. Here we develop a classification system of multiple annihilation behaviors of the equilibrium points for irregular-shaped minor celestial bodies. Most minor bodies have five equilibrium points; there are twice the number of annihilations per equilibrium point when the number of equilibrium points is between one and five. We present the detailed annihilation classifications for equilibria of objects which have five equilibrium points. Additionally, the annihilation classification for the seven equilibria of Kleopatra-shaped objects and the nine equilibria of Bennu-shaped objects are also discussed. Equilibria of different objects fall into different annihilation classifications. By letting the rotational speeds vary, we studied the annihilations and creations of relative equilibria in the gravitational field of ten minor bodies, including eight asteroids, one satellite of a planet, and one cometary nucleus: the asteroids were 216 Kleopatra, 243 Ida, 951 Gaspra, 1620 Geographos, 2063 Bacchus, 2867 Steins, 6489 Golevka, and 101955 Bennu; the satellite of the planet was S16 Prometheus; and the comet was 1682 Q1/Halley. For the asteroid 101955 Bennu, which has the largest number of equilibria among the known asteroids, we find that the equilibrium points with different indices approach each other as the rotational speed varies and thus annihilate each other successively. The equilibrium in the gravitational field and the smooth surface equilibrium collides when the equilibrium touches the surface of the body.

## 1. Introduction

Recently, it has been found that the YORP effect (Yarkovsky–O'Keefe–Radzievskii–Paddack effect) is one of the most important influences on asteroid dynamics (Scheeres, 2007; Taylor et al., 2007; Lowry et al., 2014; Ševeček et al., 2015). YORP effect is caused by the effect of the electromagnetic radiation to the surface of the asteroid. The photons of the electromagnetic radiation from the Sun has momentum, the effect makes the variety of the angular momentum of the asteroid relative to the inertial space. Although the changes are very small, when the time is long enough, the changes of the angular momentum of the body may be very significant. An example is the asteroid 54509 YORP (2000 PH5). The YORP effect causes its rotational speed to increase, and the change in spin rate is $(2.0 \pm 0.2) \times 10^{-4}$ deg·day$^{-2}$ (Taylor et al., 2007). Different asteroids have different YORP effects, the characteristic and the value of YORP effects depend on the size and irregular shape of the asteroids (Vokrouhlický and Čapek, 2002; Micheli and Paolicchi, 2008). Vokrouhlický and Čapek (2002) classified several types of YORP effects for asteroids; for instance, the YORP effect on Toutatis-shaped asteroids, including 433 Eros, 1998KY26, and 25143 Itokawa, belong to type I.

The YORP effect can lead to variations in the rotational speed, which causes the equilibrium shapes and equilibrium points of the asteroid to vary (Cotto-Figueroa et al., 2014; Hirabayashi and Scheeres, 2014). The relative equilibrium point is the point where the external force of the gravitational force and the Coriolis force equals zero. The gravitational force is caused by the irregular shape and mass distribution of the asteroid, while the Coriolis force is caused by the rotation of the asteroid. The changes in equilibrium shapes and equilibrium points influence the surface grain motion as well as the origin of asteroid pairs and triple asteroids (Walsh et al., 2012; Hirabayashi et al., 2015; Jiang et al., 2016b). To understand the characteristics of equilibrium shapes and equilibrium points, several important analytical and numerical results for conditions, stability, and classifications of equilibrium have been investigated.

Holsapple (2004) discussed the stability of equilibrium shapes by modeling the Solar System bodies as elastic-plastic solids. Richardson et al. (2005) investigated the shape and spin limits of rubble-pile Solar System bodies and found that the rubble-pile body consisting of a small number of particles is more stable than one consisting of a large number of particles. Sharma et al. (2009) produced the theoretical and numerical analysis of equilibrium shapes of gravitating rubble asteroids. Walsh et al. (2012) studied the disruption of asteroids and the formation of binary asteroids when the rubble-pile asteroids are spun up by the YORP effect. Hirabayashi and Scheeres (2014) calculated the surface shedding of the asteroid 216 Kleopatra and found that when the rotational speed changes from 5.385h to 2.81h, the outer equilibrium points will attach to the surface of the asteroid. This is the key influence on surface shedding and leads to the first-shedding condition, which happens because the radial acceleration is then oriented outwards (Scheeres et al., 2016).

To study the position, stability, and topological classification of equilibrium points, several authors have considered simple models to represent the stability and motion around equilibrium points. To understand the motion around equilibria near the elongated celestial bodies, Riaguas et al. (2001) investigated the nonlinear stability of the equilibria in the gravitational potential of a finite straight segment. Vasilkova (2005) used a triaxial ellipsoid to model the asteroid's shape and discussed the periodic motion around the equilibrium points of the triaxial ellipsoid. Palacián et al. (2006) calculated four equilibrium points in the potential of the finite straight segment and the invariant manifold near the equilibrium points. Guirao et al. (2011) discussed the position and stability of equilibria in a double-bar rotating system with changing parameters. However, a minor celestial body's irregular shape and gravitation are different from the simple models. The polyhedral model (Werner, 1994; Werner and Scheeres, 1996), which has sufficient faces to model the irregular shape and gravitation, is much more precise than the simple models. Jiang et al. (2014) presented the theory of local dynamics around equilibrium points and classified the non-degenerate equilibrium points into eight different cases. They used the polyhedral model and applied the theory to calculate the stability and topological cases around four different asteroids. Wang et al. (2014) used the classification presented in Jiang et al. (2014) and calculated the stability and topological classifications of equilibrium points around 23 minor celestial bodies; the shapes and gravitation models of these objects were also computed using the polyhedral model. Chanut et al. (2015) used the classification presented in Jiang et al. (2014) and found that there are one unstable equilibrium point and two linearly stable equilibrium points inside the asteroid 216 Kleopatra. Jiang et al. (2015) found that the equilibrium points in the potential of asteroid 216 Kleopatra may collide and annihilate each other as the rotational speed varies.

This paper investigates the annihilation and creation of equilibrium points of minor bodies when their rotation speeds vary. We take into consideration that different minor bodies may have different types of YORP effects, and different YORP effects lead to different changes in rotational speeds (Vokrouhlický and Čapek, 2002). We mainly want to calculate the annihilation classification, the kinds of bifurcations, and the annihilation positions of asteroids during changes in rotational speed. Therefore, we chose minor bodies with different YORP types, including 216 Kleopatra, 951 Gaspra, and 2063 Bacchus, etc. Asteroid 216 Kleopatra's YORP type is Type 2, 951 Gaspra's YORP type is Type 1, and 2063 Bacchus' YORP type is Type 4 (Micheli and Paolicchi, 2008). In addition, the dynamical behaviors of satellites of planets and cometary nuclei are different from those of asteroids, so we chose the satellite of planet S16 Prometheus and the comet 1682 Q1 Halley to analyze the annihilation and creation of equilibrium points around them when the rotational speeds vary.

The topological classifications and the indices of equilibrium points are analyzed. Then we present the annihilation classifications of relative equilibrium points as the parameters of the gravitational field vary. We discuss all the annihilation classifications for equilibria of objects that have five equilibrium points because most of the minor bodies have five equilibrium points. However, the Kleopatra-shaped objects may have seven equilibrium points as the rotational speed varies, and the Bennu-shaped objects may have nine equilibrium points as the rotational speed varies; therefore, we studied the annihilation classifications for seven equilibria of Kleopatra-shaped objects and nine equilibria of Bennu-shaped objects. We investigated ten objects, including the eight asteroids 216 Kleopatra, 243 Ida, 951 Gaspra, 1620 Geographos, 2063 Bacchus, 2867 Steins, 6489 Golevka, and 101955 Bennu, the satellite of planet S16 Prometheus, and the comet 1682 Q1/Halley. For asteroid 101955 Bennu, the results show that the equilibrium points with different indices approach each other when the rotational speed increases. The equilibrium points annihilate each other successively, and there are a total of four pairs of annihilations around asteroid 101955 Bennu.

The number of equilibrium points for almost axisymmetric bodies is quite variable, and elongated bodies can also have a different number of equilibrium points than those presented if there is some heterogeneity in the mass distribution. Although the number of equilibrium points for different shapes of celestial bodies seems to not have much value from the standpoint of mathematics, this phenomenological study has a wider significance. From this phenomenological study, one can summarize the characteristics of the equilibrium points, the kinds of bifurcations, as well as the annihilation positions, which are then helpful for a theoretical study.

**2. Topological Classifications and Indices of Equilibria**

In a minor body's gravitational potential the gradient of the effective potential at the relative position $\mathbf{r} = (x, y, z)^{\mathrm{T}}$ can be written as Equation (1):

$$\begin{cases} \dfrac{\partial V(\mathbf{r})}{\partial x} = -\omega^2 x + \dfrac{\partial U(\mathbf{r})}{\partial x} \\ \dfrac{\partial V(\mathbf{r})}{\partial y} = -\omega^2 y + \dfrac{\partial U(\mathbf{r})}{\partial y} \\ \dfrac{\partial V(\mathbf{r})}{\partial z} = \dfrac{\partial U(\mathbf{r})}{\partial z} \end{cases}, \qquad (1)$$

where $\mathbf{r}$ represents the relative position in the body-fixed frame, $\boldsymbol{\omega}$ represents the body's rotational angular velocity, $\omega$ represents the norm of $\boldsymbol{\omega}$, $U(\mathbf{r})$ represents the body's gravitational potential. The equilibrium point is defined as the zero point of the gradient of the effective potential (Jiang et al., 2014; Chanut et al., 2014). With its current rotation speed asteroid 1998 $KY_{26}$ is found to have only one equilibrium point (Wang et al., 2014), 216 Kleopatra has seven equilibrium points (Hirabayashi and Scheeres, 2014; Wang et al., 2014; Chanut et al., 2015), 101955 Bennu has nine equilibrium points (Wang et al., 2014); other minor celestial bodies, including asteroids 4 Vesta, 243 Ida, 433 Eros, 951 Gaspra, 1620 Geographos, 1996 HW1, 2063 Bacchus, 2867 Steins, 4769 Castalia, 6489 Golevka, 25143 Itokawa and 52760 1998 $ML_{14}$, and satellites of planets J5 Amalthea, M1 Phobos, N8 Proteus, S9 Phoebe, and S16 Prometheus, as well as comets 1P/Halley, 9P/Tempel1, and 103P/Hartley2, all have five equilibrium points (Jiang et al., 2014; Wang et al., 2014). Most of these minor celestial bodies have one equilibrium point inside the body and the other equilibrium points outside except for 216 Kleopatra. 216 Kleopatra has three equilibrium points inside (Wang et al., 2014; Chanut et al., 2015) and four equilibrium points outside (Jiang et al., 2014).

The equation of motion relative to the equilibrium point (Jiang et al., 2014; 2015a) can be expressed in the tangent space as Equation (2):

$$\mathbf{M}\ddot{\boldsymbol{\rho}} + \mathbf{G}\dot{\boldsymbol{\rho}} + (\nabla^2 V)\boldsymbol{\rho} = 0, \qquad (2)$$

where $\boldsymbol{\rho}$ represents the relative position from the equilibrium point to any point in the vicinity of the equilibrium point, $\mathbf{M}$ represents a $3\times 3$ unit matrix,

$$\mathbf{G} = \begin{pmatrix} 0 & -2\omega & 0 \\ 2\omega & 0 & 0 \\ 0 & 0 & 0 \end{pmatrix},$$ and $\nabla^2 V$ represents the effective potential's Hessian matrix.

The roots of Eq. (2) are eigenvalues of the equilibrium point. We use $E_i$ to represent the *i*th equilibrium point and assume that there are a total of $N$ equilibrium points in an asteroid's gravitational potential. Let $\lambda_j(E_i)$ be the *j*th eigenvalue of $E_i$, then for all the equilibrium points of this asteroid, one has the relation

$$\sum_{i=1}^{N} \left[ \text{sgn} \prod_{j=1}^{6} \lambda_j(E_i) \right] = const .$$

The topological classification of equilibrium points are based on the distribution of six eigenvalues on the complex plane. We use the classification definition from Jiang et al. (2016a). The most common topological classifications of non-degenerate equilibrium points (Jiang et al., 2014; Wang et al., 2014; Chanut et al., 2015) include Case O1: $\pm i\beta_j \left(\beta_j \in \mathrm{R}, \beta_j > 0; j=1,2,3\right)$; Case O2: $\pm \alpha_j \left(\alpha_j \in \mathrm{R}, \alpha_j > 0, j=1\right)$ and $\pm i\beta_j \left(\beta_j \in \mathrm{R}, \beta_j > 0; j=1,2\right)$; as well as Case O5: $\pm i\beta_j \left(\beta_j \in \mathrm{R}, \beta_j > 0, j=1\right)$ and $\pm \sigma \pm i\tau \left(\sigma, \tau \in \mathrm{R}; \sigma, \tau > 0\right)$. Equilibrium points which belong to Case O1 are linearly stable while equilibrium points which belong to Case O2 or Case O5 are unstable. The linearly stable equilibrium points have only the one topological case, i.e. Case O1 (Jiang et al., 2014; Chanut et al., 2015). The most common topological classifications of degenerate equilibrium points include Case DE2: $\gamma_j \left(\gamma_j = 0; j=1,2\right)$ and $\pm i\beta_j \left(\beta_j \in \mathrm{R}, \beta_j > 0; j=1,2 | \beta_1 \neq \beta_2\right)$ as well as Case DE5: $\gamma_j \left(\gamma_j = 0; j=1,2,3,4\right)$ and $\pm i\beta_1 \left(\beta_1 \in \mathrm{R}, \beta_1 > 0\right)$. The topological case and case distribution of the outside equilibrium points of asteroids 216 Kleopatra, 1620 Geographos, and 4769 Castalia are the same. All of them have four equilibrium points outside the body, two of them belong to Case O2 and two belong to Case O5. Different topological cases of equilibrium points have a staggered distribution, which means the topological cases of the minor celestial bodies are showed to be Case O2, Case O5, Case O2, and Case O5; or Case O2, Case O1, Case O2, and Case O1 (Wang et al., 2014). Asteroids 4 Vesta, 2867 Steins, 6489 Golevka and 52760 1998 ML$_{14}$, satellites of planets M1 Phobos, N8 Proteus, and S9 Phoebe, as well as comets 1P/Halley and 9P/Templel 1 have the same topological case and case distribution of equilibrium points. All of them have one equilibrium point inside and four equilibrium points outside; two of the outside equilibrium points belong to Case O1, and the other two outside equilibrium points belong to Case O2 (Wang et al., 2014).

Denote the index of equilibrium point $E_i$ as $ind(E_i) = \text{sgn} \prod_{j=1}^{6} \lambda_j (E_i)$, then if $E_i$ belongs to Case O1 or Case O5, $ind(E_i) = +1$; if $E_i$ belongs to Case O2, $ind(E_i) = -1$; else if $E_i$ is a degenerate equilibrium point, $ind(E_i) = 0$. Then to see the positions and indices of equilibrium points anticlockwise, we find that the equilibrium points with different indices have a staggered distribution. For instance, the asteroid 2867 Steins has five equilibrium points: $E$1, $E$2, $E$3, $E$4 are near the +$x$, +$y$, -$x$, -$y$ axes, respectively; $E$5 is near the mass center of the asteroid. The contour plot of the zero-velocity curves in a fixed plane can be calculated from the effective potential. Figure 1 shows the contour plot of the zero-velocity curve in the $xy$ plane around asteroid 2867 Steins. We can see the gradient structure of the zero-velocity curve and the position between the zero-velocity curve and the equilibrium points. Table 1 shows the eigenvalues and topological classifications of equilibrium points in the potential of 2867 Steins; the unit of the eigenvalues is s$^{-1}$. From Table 1 one can see that the topological cases of outside equilibrium points $E$1-$E$4 are O2, O1, O2, O5, respectively. This implies that the topological cases of outside equilibrium points of asteroid 2867 Steins do not have a staggered distribution.

Table 1. Eigenvalues and topological classifications of equilibrium points in the potential of 2867 Steins

| $\times 10^{-3}$s$^{-1}$ | Cases | $\lambda_1$ | $\lambda_2$ | $\lambda_3$ | $\lambda_4$ | $\lambda_5$ | $\lambda_6$ |
|---|---|---|---|---|---|---|---|
| E1 | O2 | 0.30339i | -0.30339i | 0.29342i | -0.29342i | 0.10814 | -0.10814 |
| E2 | O1 | 0.29850i | -0.29850i | 0.25548i | -0.25548i | 0.10989i | -0.10989i |
| E3 | O2 | 0.31023i | -0.31023i | 0.29656i | -0.29656i | 0.13320 | -0.13320 |
| E4 | O5 | 0.29737i | -0.29737i | 0.00359+0.19754i | 0.00359-0.19754i | -0.00359+0.19754i | -0.00359-0.19754i |
| E5 | O1 | 1.18948i | -1.18948i | 1.11423i | -1.11423i | 0.60141i | -0.60141i |

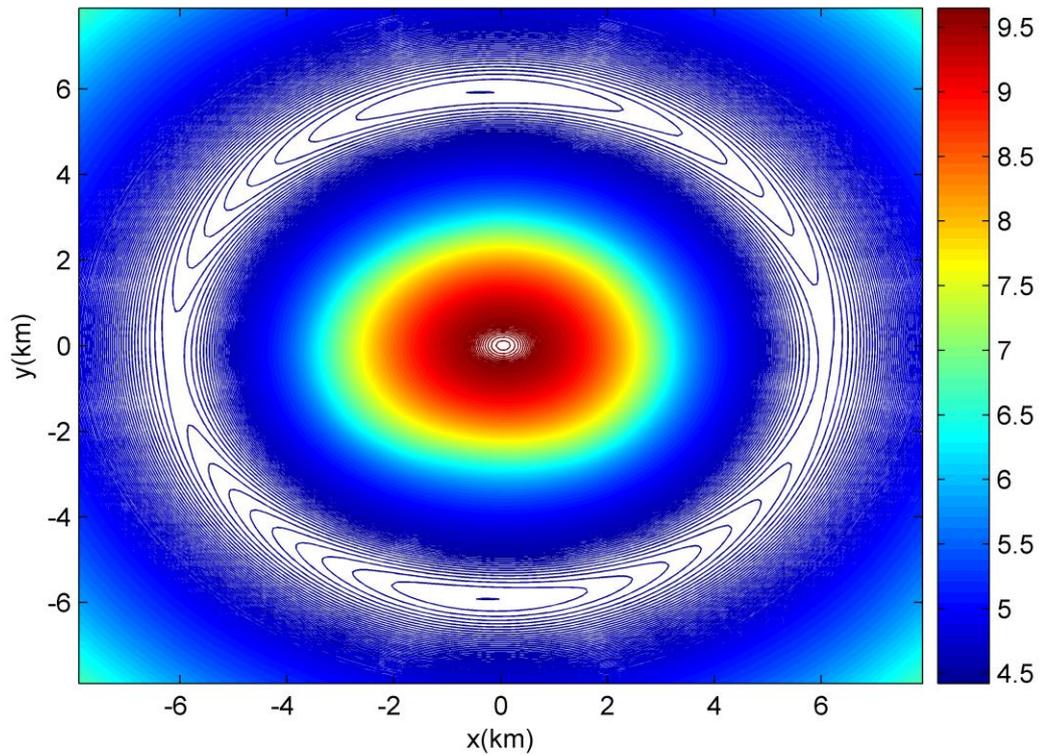

Figure 1. The contour plot of the zero-velocity curves in the *xy* plane around asteroid 2867 Steins, the unit of the value is m$^2 \cdot$s$^{-2}$.

## 3. Annihilations

Yu and Baoyin (2012) calculated the location of equilibrium points for asteroid 216 Kleopatra. Hirabayashi and Scheeres (2014) also calculated the location of equilibrium points for asteroid 216 Kleopatra and compared their results with those in Yu and Baoyin (2012). They also calculated the change in location of equilibrium points for 216 Kleopatra. However, they did not study the stability, topological cases, and the kinds of bifurcation that occur when the equilibrium points collided. Wang et al. (2014) presented the location, stability, and topological cases of equilibrium points for 101955 Bennu and several other minor bodies. Jiang et al. (2015) investigated the change in location of equilibrium points for 216 Kleopatra, and analyzed the stability, topological cases, and the kinds of bifurcation when the equilibrium points collided and annihilated. Their paper did not present the contour plot of the effective potential or the locations of equilibrium points during the collision and annihilation. Jiang et al. (2016) investigated the change in location of equilibrium points for 216 Kleopatra, 2063 Bacchus, as well as 25143 Itokawa. However, the aim of their paper was not to investigate the collision of equilibrium points. They studied the Hopf bifurcation and the stability change when the Hopf bifurcation occurs. Scheeres (2016) presented the change in location of equilibrium points for 101955 Bennu during the change of density.

To study the whole span of spin rates to find the maximum number of equilibria is useful for defining the gravitational field (Wang et al., 2014), the surface shedding (Yu and Michel, 2018), and the structure stability (Yu et al., 2017) of minor celestial bodies. Suppose the rotational speed increases, then bifurcations of equilibrium points occur, and the equilibrium points will collide and annihilate each other. Finally, there will be only one equilibrium point left in the gravitational field of the asteroid. The collision and annihilation of two different equilibrium points include several different situations. The two most common collisional annihilations lead to the Saddle-Node bifurcation and the Saddle-Saddle bifurcation (Jiang et al., 2015). In the Saddle-Node bifurcation, two equilibrium points belonging to Case O1 and Case O2 collide and coincide to a degenerate equilibrium which belongs to Case DE2; in the Saddle-Saddle bifurcation, two equilibrium points belonging to Case O2 and Case O5 collide and coincide to a degenerate equilibrium which belongs to Case DE5. In both of these two bifurcations the degenerate equilibrium will disappear as the rotational speed continues to vary. The creation of equilibrium points is opposite to the annihilation of them; for instance, the creation which leads to the Saddle-Node bifurcation has a degenerate equilibrium point which belongs to Case DE2 which creates and separates into two non-degenerate equilibrium points belonging to Case O1 and Case O2.

Most of the irregular-shaped minor celestial bodies, including asteroids, satellites of planets, and comets, have five equilibrium points. So we first discuss the annihilation classification for five equilibrium points.

## 3.1 Annihilation Classifications for Five Equilibria of Irregular-shaped Minor Celestial Bodies

We now consider the distribution of topological classifications of equilibrium points around the asteroids which have five equilibrium points. Most asteroids have five equilibrium points and the equilibrium points belong to Case O1, Case O2, or Case O5 (Jiang et al., 2014; Wang et al., 2014); because of the formula $\sum_{i=1}^{N} ind(E_i) = +1$ (Jiang et al., 2015), the sum of the indices of the other four equilibrium points is zero. For the asteroids which have five equilibrium points these four equilibrium points are outside the body of the asteroids, and the stable equilibrium point is inside the body of the asteroids. The topological classifications of the other four equilibrium points fall into three topological classifications: **TC1)** two of them belong to Case O1 and the other two belong to Case O2; **TC2)** two of them belong to Case O2 and the other two belong to Case O5; **TC3)** two of them belong to Case O2, one belongs to Case O1 and the remaining one belongs to Case O5. During the annihilations of two equilibrium points, only one Case O1 equilibrium point and one Case O2 equilibrium point can collide and annihilate each other, or one Case O2 equilibrium point and one Case O5 equilibrium point can collide and annihilate each other. Equilibrium points belonging to the same topological classification cannot collide and annihilate each other. Besides, it is also impossible for one Case O1 equilibrium point and one Case O5 equilibrium point to collide and annihilate each other. If the topological classification of the outside four equilibrium points is TC1 or TC2, equilibrium points that belong to different topological cases will have a staggered distribution such that two adjacent equilibrium points can collide and annihilate each other as the rotational speed varies. If the topological classification of the outside four equilibrium points is TC3, equilibrium points that belong to different indices will have a staggered distribution such that two adjacent equilibrium points can collide and annihilate each other as the rotational speed varies. For this situation consider the anticlockwise order of the four outside equilibrium points, the four equilibrium points belong to Case O1, Case O2, Case O5, and Case O2, respectively. Similarly consider the order of the four outside equilibrium points for TC1, the order of topological classifications of equilibrium points are Case O1, Case O2, Case O1, and Case O2; for TC2, the order of topological classifications of equilibrium points are Case O2, Case O5, Case O2, and Case O5.

The innate character for distributions of equilibrium points outside the body of asteroids is that equilibrium points belonging to different indices have a staggered distribution such that two adjacent equilibrium points can collide and annihilate each other as the rotational speed varies (Jiang et al., 2015). Now we discuss the annihilation classification for equilibrium points in the gravitational field of irregular-shaped minor celestial bodies as the rotational speed changes.

The body-fixed frame is defined such that the $x$, $y$, and $z$ axes are the smallest, middle, and biggest inertia moment, respectively; and the origin of the coordinate system is the mass centre of the asteroid. Because of the irregular shape of the asteroid, in the asteroid's body-fixed frame, all the equilibrium points are out-of-plane equilibria. However, the components of the positions of equilibrium points on the *z*-axis are always small, and most of the outside equilibrium points are near the *x* axis and *y* axis. For those asteroids which have five equilibrium points with four of them outside and one inside, only the equilibrium point inside the body always remains near the origin of the body-fixed frame. We present the annihilation classifications with figures. There are seven types of annihilation classifications for five equilibrium points.

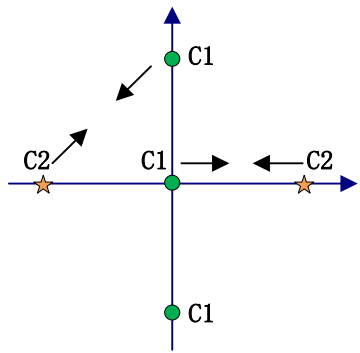

Type I

(a)

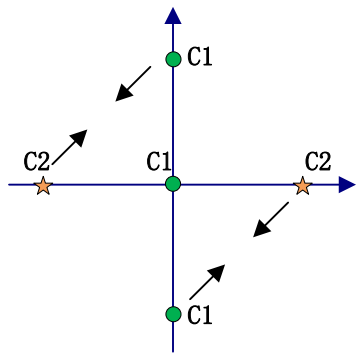

Type II

(b)

Figure 2 Annihilation classification for five equilibrium points when topological classifications of equilibrium points belong to TC1, circle dots represent equilibrium points which belong to Case O1 while stars represent equilibrium points which belong to Case O2.

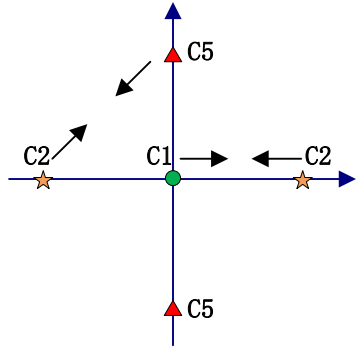

Type III

(a)

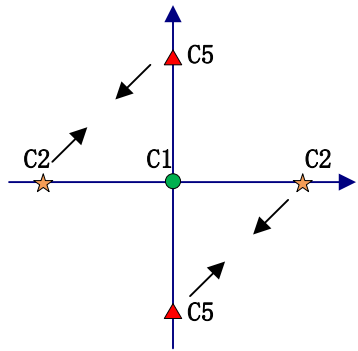

Type IV

(b)

Figure 3 Annihilation classification for five equilibrium points when topological classifications of equilibrium points belong to TC2, circle dots represent equilibrium points which belong to Case O1, stars represent equilibrium points which belong to Case O2, and triangles represent equilibrium points which belong to Case O5.

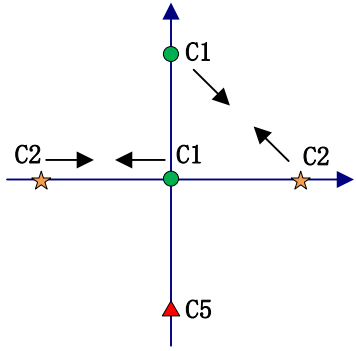

Type V

(a)

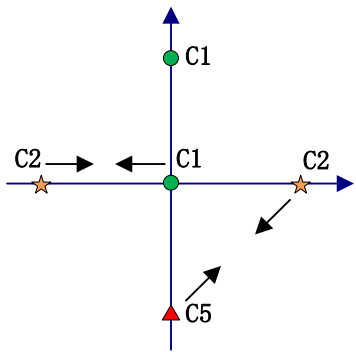

Type VI

(b)

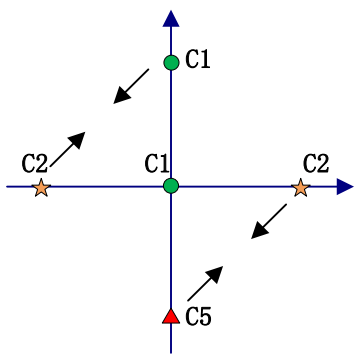

Type VII

(c)

Figure4 Annihilation classification for five equilibrium points when topological classifications of equilibrium points belong to TC3, circle dots represent equilibrium points which belong to Case

O1, stars represent equilibrium points which belong to Case O2, and triangles represent equilibrium points which belong to Case O5.

When the topological classifications of the equilibrium points belong to TC1, there are two types of annihilation classifications which are presented in Figure 2, i.e. Type I and Type II. We use circle dots to represent Case O1 equilibrium points, stars to represent Case O2 equilibrium points, and triangles to represent Case O5 equilibrium points. The abscissa represents the $x$ axis of the asteroid's body fixed frame while the ordinate represents the $y$ axis of the asteroid's body fixed frame. The dots, stars, and triangles on the axes represent the equilibrium points near the axes of asteroid's body fixed frame.

Let the rotational speed vary. In Type I the centre equilibrium point which belongs to Case O1 and the $x$ axis equilibrium point will approach each other, and then collide and annihilate each other. The $x$ axis equilibrium point and the $y$ axis equilibrium point will annihilate each other. In Type II, the centre equilibrium point which belongs to Case O1 will not collide and annihilate with the other equilibrium points. The $x$ axis equilibrium point and the $y$ axis equilibrium point will annihilate each other.

When the topological classifications of equilibrium points belong to TC2, there are two types of annihilation classifications which are presented in Figure 3, i.e. Type III and Type IV. In Type III-a the centre equilibrium point which belongs to Case O1 and the $x$ axis equilibrium point which belongs to Case O2 will annihilate each other.

The *x* axis equilibrium point which belongs to Case O2 and the *y* axis equilibrium point which belongs to Case O5 will annihilate each other. The other one of *y* axis equilibrium points which belongs to Case O5 will be left and not collide and annihilate with other equilibrium points. Although for the five-point equilibria type I and type III look similar, they are actually different. For type I the two annihilations all belong to the Saddle-Node bifurcation; however for type III one belongs to the Saddle-Node bifurcation and the other belongs to the Saddle-Saddle bifurcation. In Type IV the centre equilibrium point which belongs to Case O1 will never annihilate with other equilibrium points. The *x* axis and the *y* axis equilibrium points which belong to Case O2 and Case O5, respectively will annihilate each other.

When the topological classifications of equilibrium points belong to TC3, there are three types of annihilation classifications which are presented in Figure 4, i.e. Type V, Type VI, and Type VII. In Type V the centre equilibrium point which belongs to Case O1 and the *x* axis equilibrium point which belongs to Case O2 will annihilate each other. The *x* axis and the *y* axis equilibrium points which belong to Case O2 and Case O1, respectively will annihilate each other. The *y* axis equilibrium point which belongs to Case O5 will never annihilate other equilibrium points.

In Type VI the centre equilibrium point which belongs to Case O1 and the *x* axis equilibrium point which belongs to Case O2 will annihilate each other. In addition, the *x* axis and the *y* axis equilibrium points will annihilate each other, which belong to Case O2 and Case O5, respectively. The *y* axis equilibrium point which belong to Case O1 will be left.

In Type VII the *x* axis and the *y* axis equilibrium points which belong to Case O2 and Case O5, respectively will annihilate each other,; and the *x* axis and the *y* axis equilibrium points which belong to Case O2 and Case O1, respectively will annihilate each other,. The centre equilibrium point will be left.

We now analyze the annihilation classifications for several objects which have five equilibrium points, including asteroids 243 Ida, 951 Gaspra, 1620 Geographos, 2867 Steins, 3103 Eger, and the satellite of planet S16 Prometheus. The physical parameters we used are listed in Table A1 in Appendix A.

### 3.1.1  243 Ida

The annihilation classification of asteroid 243 Ida belongs to Type III with five equilibrium points. Figure 5 shows the annihilation of the five equilibrium points of asteroid 243 Ida as the rotational speed varies. The positions of annihilation of equilibria around the body and kinds of bifurcations are listed in Table A2 in Appendix A.

The first annihilation, *E*1, moves to the center of the body, and *E*1 and *E*5 collide and annihilate each other inside the body of the asteroid. *E*1 belongs to Case O2 while *E*5 belongs to Case O1. During the first annihilation the Saddle-Node bifurcation occurs. Topological cases of *E*2, *E*3, and *E*4 remain unchanged after the annihilation of *E*1 and *E*5. After the first annihilation no linearly stable equilibrium points exist.

In the second annihilation *E*3 and *E*4 collide on the surface of the asteroid. During the first annihilation the Saddle-Saddle bifurcation occurs. The topological case for equilibrium point *E*2 changes from Case O5 to Case O1 before the second annihilation. After the second annihilation only the equilibrium point *E*2 is left.

As the rotational speed varies, *E*2 enters the body, although it is of little use to astrodynamics anymore, it still retains some value for the scientific research of the structural dynamics of the minor bodies. Although minor celestial bodies are likely to be heterogeneous at the small scale, most of them have a density almost equal throughout the different parts except for a few bodies such as 25143 Itokawa (Lowry et al., 2014). Thus the study of the inner equilibrium point in a homogeneous medium body is also useful.

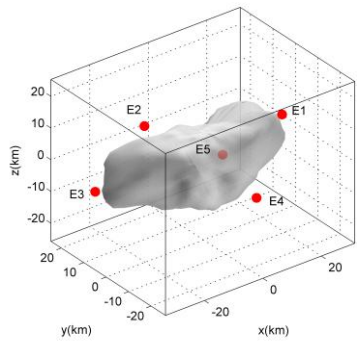 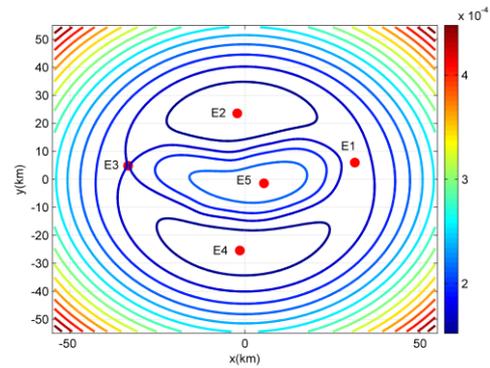

(a)

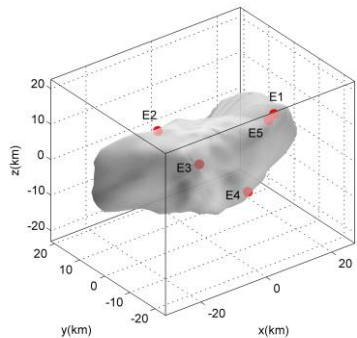 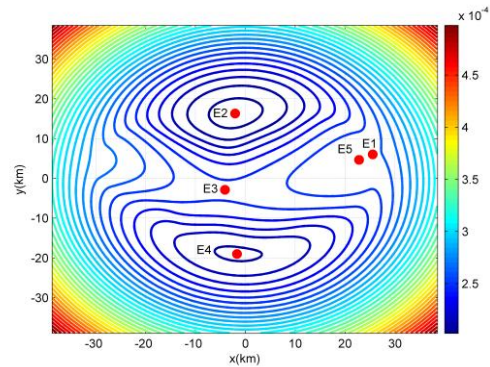

(b)

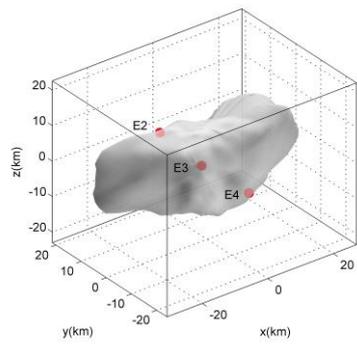 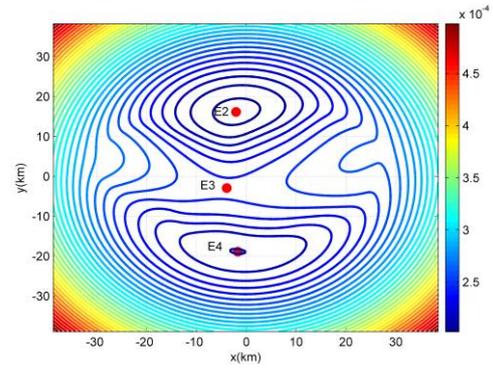

(c)

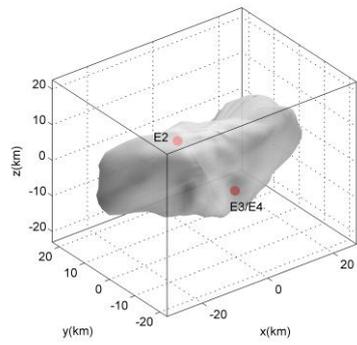 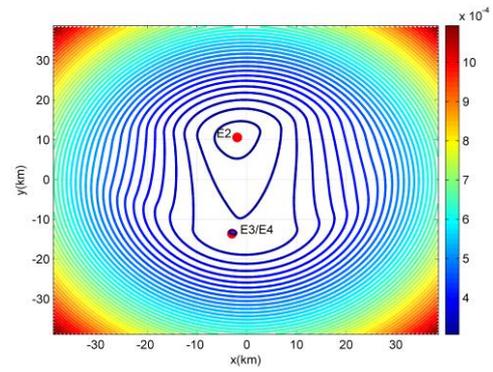

(d)

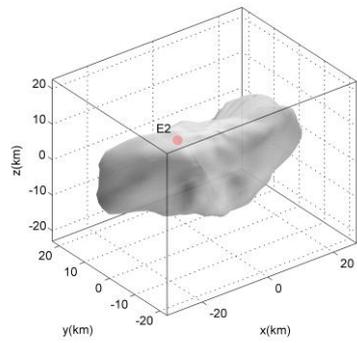 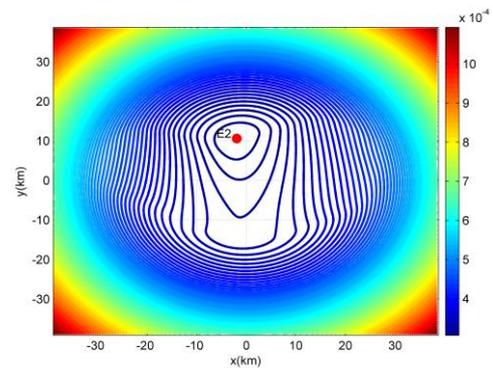

(e)

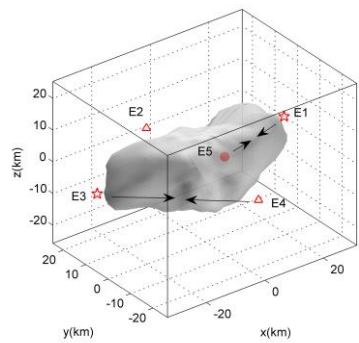

(f)

Figure5 The annihilation of five equilibrium points of asteroid 243 Ida as the rotational speed varies: (a) $\omega=1.0\omega_0$; (b) $\omega=1.4992\omega_0$; (c) $\omega=1.4993\omega_0$; (d) $\omega= 2.24719\omega_0$; (e) $\omega= 2.24720\omega_0$; (f) the variation trend of the equilibrium points.

### 3.1.2　951 Gaspra

The annihilation classification of asteroid 951 Gaspra is Type III for five equilibrium points. Figure 6 shows the annihilation of the five equilibrium points of asteroid 951 Gaspra as the rotational speed varies. Just as before the positions of annihilation of equilibria around the body and kinds of bifurcations are listed in Table A2 in Appendix A.

The first annihilation, $E1$, moves to the center of the body, and $E3$ and $E5$ collide near the intersection point caused by the $-x$ axis and the surface. During the first annihilation the Saddle-Node bifurcation occurs. $E1$ belongs to Case O2, $E3$ belongs to Case O2, and $E5$ belongs to Case O1. Topological cases of $E1$, $E2$, and $E4$ remain unchanged after the annihilation of $E3$ and $E5$. After the annihilation no linearly stable equilibrium points exist.

In the second annihilation $E1$ and $E4$ collide near the intersection point caused by the $-y$ axis and the surface. During the second annihilation the Saddle-Saddle bifurcation occurs. The topological case of equilibrium point $E2$ changes from Case O5 to Case O1 before the second annihilation.

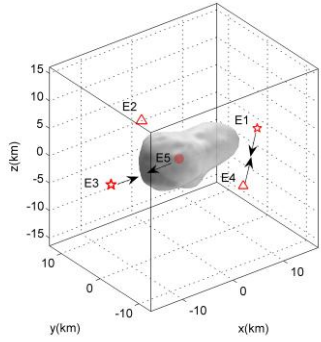

Figure 6 The annihilation of five equilibrium points of asteroid 951 Gaspra as the rotational speed varies: the variation trend of the equilibrium points.

### 3.1.3    1620 Geographos

The annihilation classification of asteroid 1620 Geographos belongs to Type III with five equilibrium points. Figure 7 shows the annihilation of the five equilibrium points of asteroid 1620 Geographos as the rotational speed varies.

In the first annihilation $E1$ and $E3$ move towards the body of the asteroid, where the direction of motion for $E5$ is along the $+x$ axis. $E1$ and $E5$ collide inside the body of the asteroid. The Saddle-Node bifurcation occurs in this annihilation. During the first annihilation the directions of motion for $E2$ and $E4$ point to the body center of the asteroid, and these two equilibrium points are still outside the body of the asteroid. The topological cases of the equilibrium points do not change before the first annihilation.

After the first annihilation there are only three equilibrium points left, which are $E2$, $E3$, and $E4$, respectively. $E3$ is inside the body while $E2$ and $E4$ are outside. As the rotational speed continues to vary, $E2$ and $E3$ approach each other, and $E4$ moves to the body of the asteroid. When $\omega = 2.365\omega_0$, $E2$ and $E3$ collide inside the body, and *E*4 is just on the surface of the asteroid. The Saddle-Saddle bifurcation occurs in the second annihilation. After the second annihilation only *E*4 remains. As the rotational speed continues to vary, *E*4 moves into the body. The topological case of equilibrium point *E*4 changes from Case O5 to Case O1 before the second annihilation.

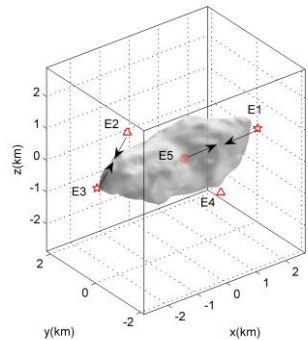

Figure 7 The annihilation of five equilibrium points of asteroid 1620 Geographos as the rotational speed varies: the variation trend of the equilibrium points.

### 3.1.4 2867 Steins

The annihilation classification of asteroid 2867 Steins belongs to Type IV with five equilibrium points. Figure 8 shows the annihilation of the five equilibrium points of asteroid 2867 Steins as the rotational speed varies. Before the first annihilation *E*1 and *E*3 move towards the body of the asteroid, the direction of motion for *E*5 is along the +*x* axis while the direction of motion for *E*1 is along the -*x* axis. The topological cases of the equilibrium points do not change before the first annihilation. *E*1 and *E*5 collide inside the asteroid; the bifurcation belongs to the Saddle-Node bifurcation. During the first annihilation the directions of motion for *E*2 and *E*4 point to the body center of the asteroid. *E*3 moves into the body while *E*2 and *E*4 remain on the surface of the asteroid when the first annihilation occurs.

After the first annihilation there are only three equilibrium points left, which are *E*2, *E*3, and *E*4, respectively. *E*3 is inside the body while *E*2 and *E*4 are on the surface of the body. *E*2 belongs to Case 5 while *E*3 belongs to Case 2. The topological case of equilibrium point *E*4 changes from Case O5 to Case O1 before the second annihilation. As the rotational speed continues to vary, *E*2 and *E*3 approach each other, and they collide inside the asteroid when $\omega = 3.26\omega_0$. The bifurcation of the second annihilation also belongs to the Saddle-Node bifurcation. When $\omega = 3.26\omega_0$, *E*4 is on the surface of the asteroid. After the second annihilation only *E*4 remains. As the rotational speed continues to vary, *E*4 moves into the body.

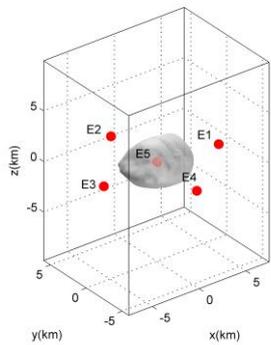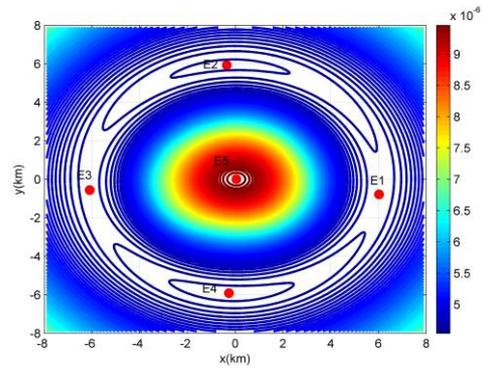

(a)

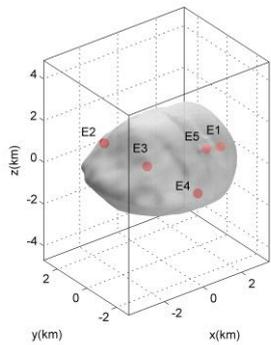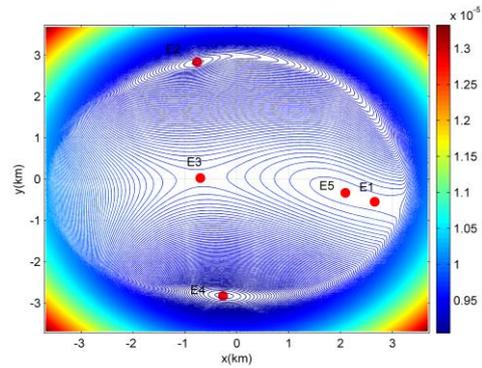

(b)

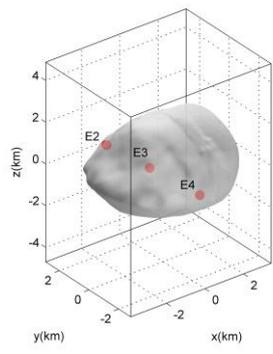 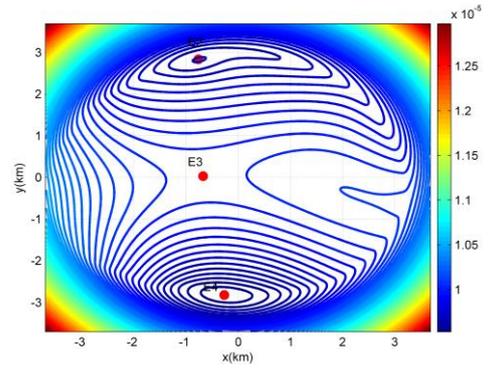

(c)

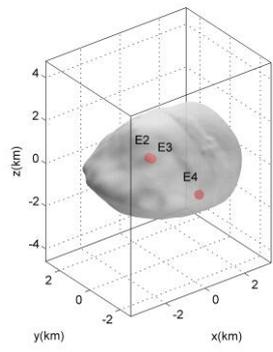 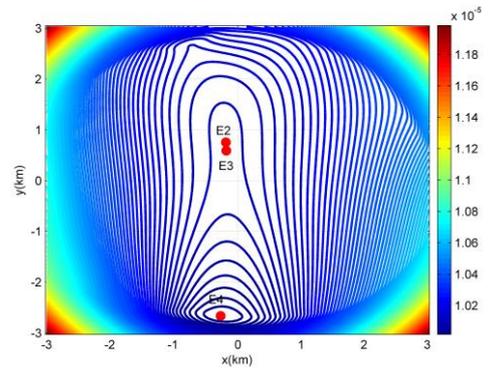

(d)

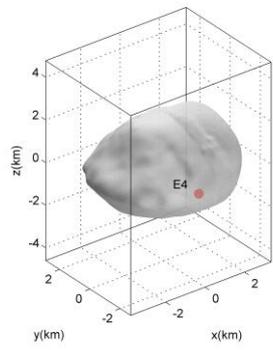 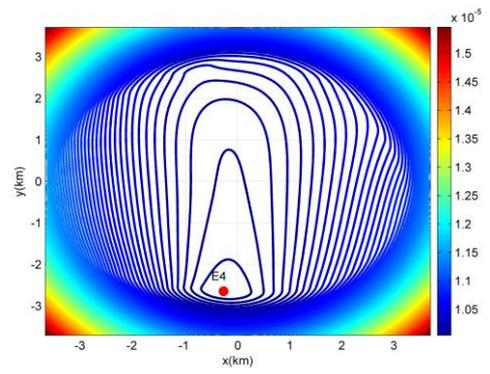

(e)

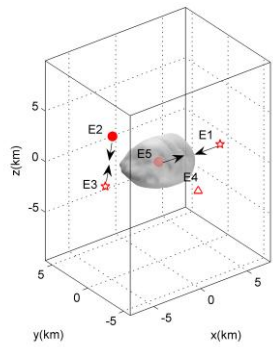

(f)

Figure 8 The annihilation of five equilibrium points of asteroid 2867 Steins as the rotational speed varies: (a) $\omega=1.0\omega_0$; (b) $\omega=2.972\omega_0$; (c) $\omega=2.976\omega_0$; (d) $\omega=3.25\omega_0$; (e) $\omega=3.28\omega_0$; (f) the variation trend of the equilibrium points.

### 3.1.5  6489 Golevka

The annihilation classification of asteroid 6489 Golevka is Type I with five equilibrium points. Figure 9 shows the annihilation of the five equilibrium points of asteroid 6489 Golevka as the rotational speed varies. Before the first annihilation all the outside equilibrium points move towards the body of the asteroid 6489 Golevka; the direction of motion for *E*5 is along the +*x* axis while the direction of motion for *E*1 is along the -*x* axis. *E*1 and *E*5 collide and annihilate each other inside the asteroid when $\omega=2.696\omega_0$, and the bifurcation of the first annihilation is a Saddle-Node bifurcation. The topological cases of equilibrium points do not change before the first annihilation.

After the first annihilation there are only three equilibrium points left, which are *E*2, *E*3, and *E*4, respectively. *E*3 is inside the body while *E*2 and *E*4 are outside. *E*2 and *E*4 belong to Case O5 while *E*3 belongs to Case O2. As the rotational speed varies from $\omega=2.696\omega_0$ to $\omega=3.437\omega_0$, *E*4 moves into the body of the asteroid. The topological case of equilibrium point *E*4 changes from Case O5 to Case O1 before the second annihilation. *E*2 and *E*3 approach each other, and they collide on the surface of the asteroid when $\omega=3.437\omega_0$. The bifurcation of the second annihilation also belongs to the Saddle-Node bifurcation. When $\omega=3.437\omega_0$, *E*4 is inside the asteroid. After the second annihilation only *E*4 is left.

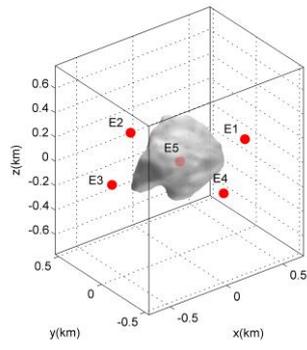 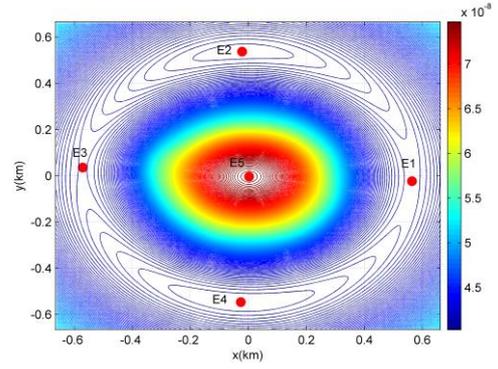

(a)

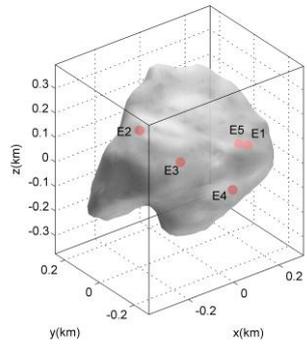 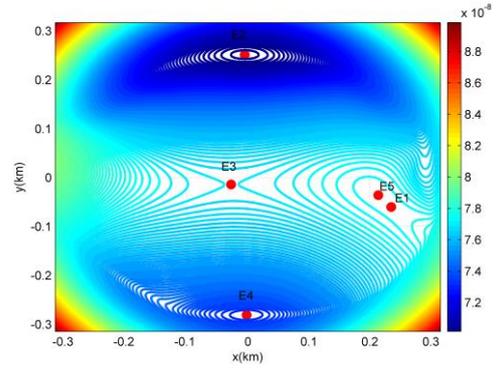

(b)

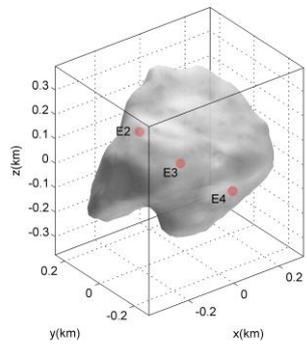 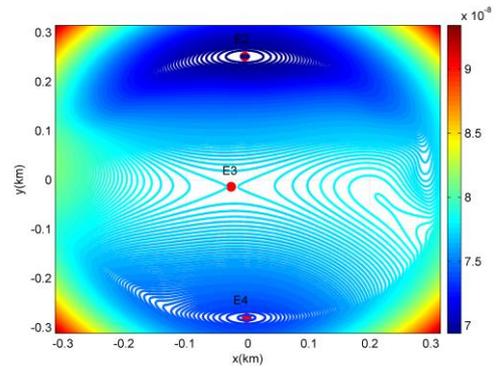

(c)

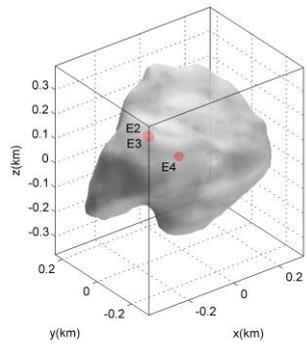 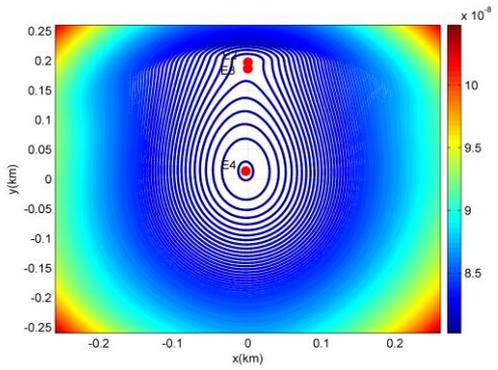

(d)

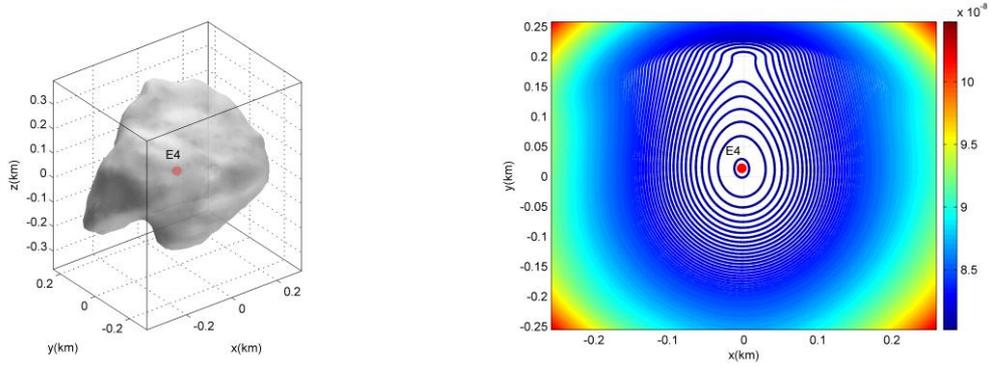

(e)

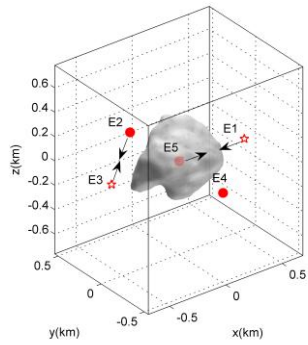

(f)

Figure 9 The annihilation of five equilibrium points of asteroid 2867 Steins as the rotational speed varies: (a) $\omega=1.0\omega_0$; (b) $\omega=2.696\omega_0$; (c) $\omega=2.699\omega_0$; (d) $\omega=3.437\omega_0$; (e) $\omega=3.442\omega_0$; (f) the variation trend of the equilibrium points.

### 3.1.6  S16 Prometheus

The annihilation classification of satellite of planet S16 Prometheus is Type III with five equilibrium points. Figure 10 shows the annihilation of the five equilibrium points of S16 Prometheus as the rotational speed varies. Before the first annihilation, all the outside equilibrium points move toward the body of S16 Prometheus, while the rotational speed changes from $\omega=\omega_0$ to $\omega=2.270\omega_0$. The topological cases of equilibrium points do not change before the first annihilation. $E3$ and $E5$ move into the body, and $E3$ collides with $E5$ inside the body of S16 Prometheus. The annihilation between $E3$ and $E5$ occurs when $\omega=2.270\omega_0$. The bifurcation of the first annihilation belongs to the Saddle-Node bifurcation.

After the first annihilation only $E$1, $E$2, and $E$4 are left. $E$1 is inside the body while $E$2 and $E$4 are on the surface. $E$2 and $E$4 belong to Case O5 while $E$1 belongs to Case O1. As the rotational speed continues to vary from $\omega = 2.27\omega_0$ to $\omega = 3.068\omega_0$, $E$2 and $E$4 move into the body of S16 Prometheus. The topological case of equilibrium point $E$4 changes from Case O5 to Case O1 before the second annihilation. $E$1 and $E$2 approach each other and collide inside the body of S16 Prometheus when $\omega = 3.068\omega_0$. The bifurcation of the second annihilation belongs to the Saddle-Saddle bifurcation. After the second annihilation only $E$4 remains. $E$4 is inside the body and belongs to Case O5.

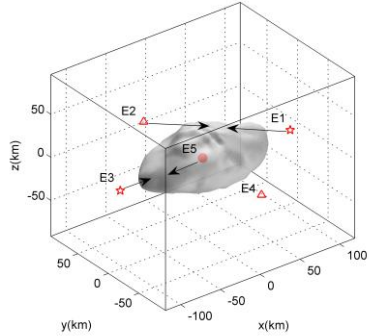

Figure 10 The annihilation of five equilibrium points of S16 Prometheus as the rotational speed varies: the variation trend of the equilibrium points.

## 3.2 Annihilation Classification for the Seven Equilibria of Kleopatra-shaped Objects and the Nine Equilibria of Bennu-shaped Objects

Asteroid 216 Kleopatra has seven equilibrium points; the two inside equilibrium points are linearly stable, and the other five equilibrium points are unstable (Jiang et al., 2014; Wang et al., 2014; Hirabayashi and Scheeres, 2014; Chanut et al., 2015).

Out of all the solar system minor bodies which have precise physical models of irregular shape, density, and rotational speed, only asteroid 216 Kleopatra is found to have seven equilibrium points. It may be easy to construct a body which has several equilibrium points; however, the annihilation of the equilibrium points may be very complicated. For the irregular-shaped bodies which have seven equilibrium points, the topological classifications of the equilibrium points have distributions that fall into several different cases; each case has its own annihilation classification. We only discuss the annihilation classification for the seven equilibria of Kleopatra-shaped objects. For the Kleopatra-shaped objects the centre equilibrium point belongs to Case O2, the $+x$ axis has two equilibrium points, one is outside the body of the object and the other one is inside the body of the object. The situation for the $-x$ axis is similar to that of the $+x$ axis. The inner one on the $+x$ axis belongs to Case O1 while the outer one belongs to Case O2. The $+y$ axis has only one equilibrium point which belongs to Case O5, the situation for the $-y$ axis is similar to the $+y$ axis.

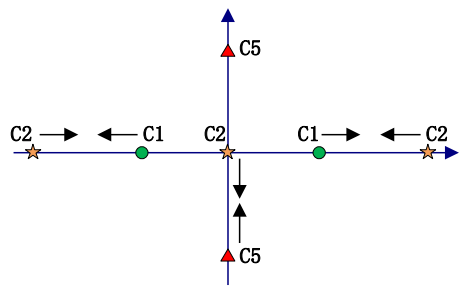

Type I

(a)

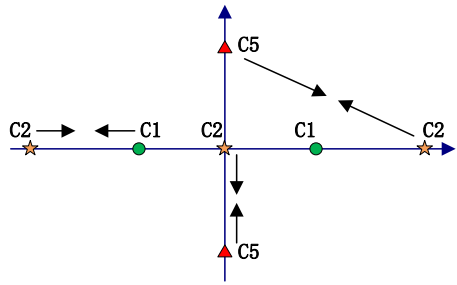

Type II

(b)

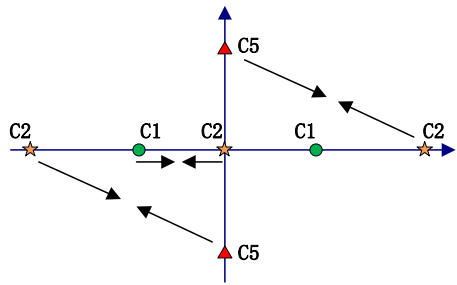

Type III

(c)

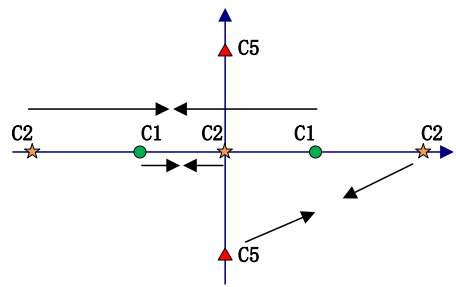

Type IV

(d)

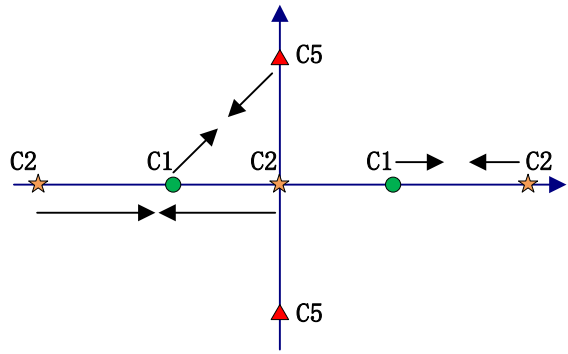

Type V

(e)

Figure 11 Annihilation classification for seven equilibrium points of Kleopatra-shaped objects

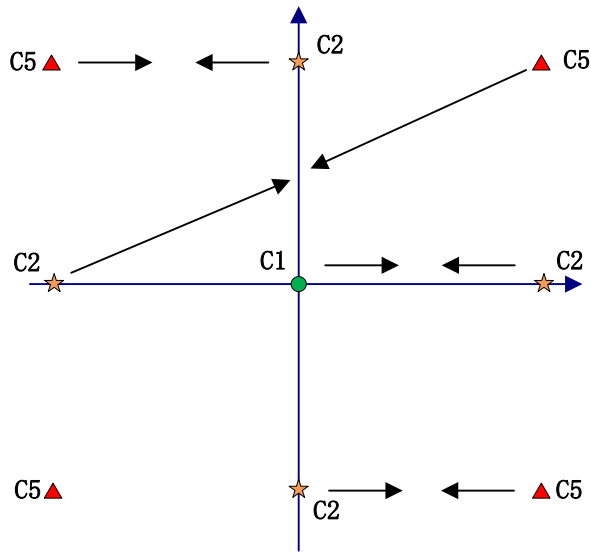

Type I

Figure 12 Annihilation classification for nine equilibrium points of Bennu-shaped objects

There are three types of annihilation classifications for the seven equilibria of Kleopatra-shaped objects, as shown in Figure 11. In Type I the centre equilibrium point which belongs to Case O2 and the $-y$ axis equilibrium point which belongs to Case O5 will annihilate each other. The two $+x$ axis equilibrium points will approach and annihilate each other, the inner one belongs to Case O1 and the outer one belongs to Case O2. The two -$x$ axis equilibrium points will also approach and annihilate each other; the inner one belongs to Case O1 and the outer one belongs to Case O2. The +$y$ axis equilibrium point will be left and will never annihilate with the other equilibrium points. The topological case of this equilibrium point will change to Case O1 before the last linearly stable equilibrium point is annihilated.

In Type II for the seven equilibria of Kleopatra-shaped objects, the centre equilibrium point and the $y$ axis equilibrium point will annihilate each other; they belong to Case O2 and Case O5, respectively. The inner and the outer equilibrium points on the $x$ axis will approach and annihilate each other; they belong to Case O1 and Case O2, respectively. The outer equilibrium point on the $x$ axis and the equilibrium point on the $y$ axis will approach and annihilate each other; they belong to Case O2 and Case O5, respectively. The inner equilibrium point on the $x$ axis will be left and will never annihilate with the other equilibrium points; the topological case of this equilibrium point remains unchanged and the position varies.

In Type III for the seven equilibria of Kleopatra-shaped objects the centre equilibrium point and the $x$ axis inner equilibrium point will annihilate each other; they belong to Case O2 and Case O1, respectively. The outer one on the $x$ axis and the one on the $y$ axis will annihilate each other; they belong to Case O2 and Case O5, respectively. The inner equilibrium point on the $x$ axis will never annihilate with the other equilibrium points, its topological case remains unchanged but its position varies.

In Type IV for the seven equilibria of Kleopatra-shaped objects, the centre equilibrium point and the *x* axis inner equilibrium point will annihilate each other, they belong to Case O2 and Case O1, respectively. The outer equilibrium point on the *x* axis and the inner equilibrium point on the *x* axis will annihilate each other; they belong to Case O2 and Case O1, respectively. The outer one on the *x* axis and the one on the *y* axis will annihilate each other; they belong to Case O2 and Case O5, respectively.

Asteroid 101955 Bennu has nine equilibrium points, only one is inside the body of the asteroid; the others are outside (Wang et al., 2014). The inside equilibrium point is linearly stable and belongs to Case O1, and the other eight equilibrium points are unstable and belong to Case O2 or Case O5. Only asteroid 101955 Bennu is found to have nine equilibrium points (Wang et al., 2014). This is the largest number of equilibrium points in the known solar system minor bodies which have a precise physical model. For the irregular-shaped bodies which have nine equilibrium points, there are several different cases for the distributions of the topological classifications for the equilibrium points, each case has its own annihilation classification. We only discuss the annihilation classification for the nine equilibria of Bennu-shaped objects.

For the Bennu-shaped objects the centre equilibrium point belongs to Case O1; going anticlockwise the topological cases of the outer eight equilibrium points are Case O2, Case O5, Case O2, Case O5, …, i.e., the topological cases have a staggered distribution (Wang et al., 2014). We only present the annihilation classification type for the nine equilibria of asteroid 101955 Bennu, as shown in Figure 12. We denote it as annihilation classification Type I for the nine equilibria of Bennu-shaped objects. In Type I the adjacent two outer equilibrium points that belong to different topological cases will approach and annihilate each other. There are a total of four collisions and annihilations. The inner equilibrium point will be left and never annihilate with the other equilibrium points; its topological case remains unchanged but its position varies.

We now analyze the annihilation classifications for several objects which have seven or nine equilibrium points, including asteroids 216 Kleopatra, 2063 Bacchus, 101955 Bennu, and the comet 1682 Q1 Halley. Only asteroid 101955 Bennu has nine equilibrium points. The physical parameters we used are listed in Table A1 in Appendix A.

### 3.2.1    216 Kleopatra

The annihilation classification of asteroid 216 Kleopatra is Type I for the seven equilibrium points of Kleopatra-shaped objects. Figure 13 shows the annihilation of five equilibrium points of 216 Kleopatra as the rotational speed varies. Before the first annihilation $E3$ and $E6$ approach each other. When the rotational speed is $\omega = 1.944586\omega_0$, $E3$ and $E6$ collide and annihilate each other on the surface of the asteroid 216 Kleopatra. The first annihilation belongs to the Saddle-Node bifurcation. After the first annihilation there are five equilibrium points left; only $E5$ is linearly stable, the others are all unstable.

As the rotational speed continues to vary $E1$ and $E5$ approach each other and touch the surface of the asteroid at the same time. These two equilibrium points touch the same point on the asteroid's surface and annihilate each other on the surface. The second annihilation also belongs to the Saddle-Node bifurcation. The topological cases of equilibrium points do not change before the first annihilation and the second annihilation. After the second annihilation there are three equilibrium points, i.e. $E$2, $E$4, and $E$7. Only $E$7 is inside the asteroid.

As the rotational speed continues to vary, $E$2 moves toward the body of the asteroid, and $E$4 and $E$7 approach each other. The topological case of equilibrium point $E$2 changes from Case O5 to Case O1 before the second annihilation. When $\omega = 4.270772\omega_0$, $E$4 and $E$7 touch the same point on the asteroid's surface; at the same time, $E$2 also touches the surface of the asteroid. After that $E$4 and $E$7 annihilate each other on the surface and $E$2 moves into the body of the asteroid. The third annihilation belongs to the Saddle-Saddle bifurcation. After the annihilation there is only one equilibrium $E$2 left.

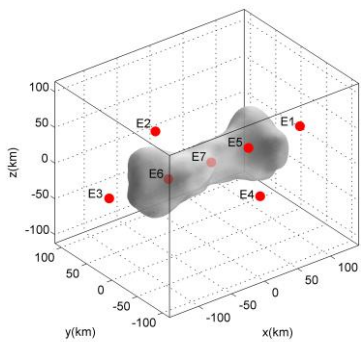 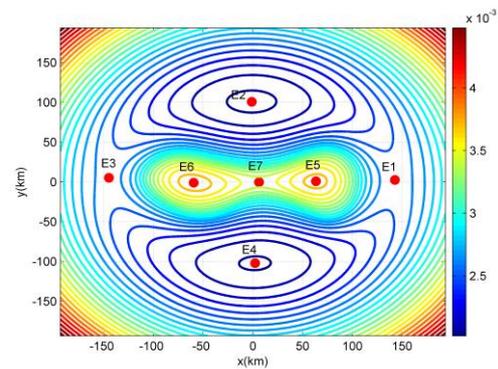

(a)

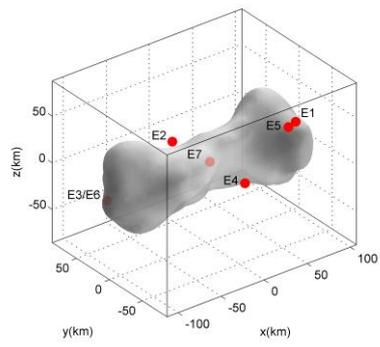 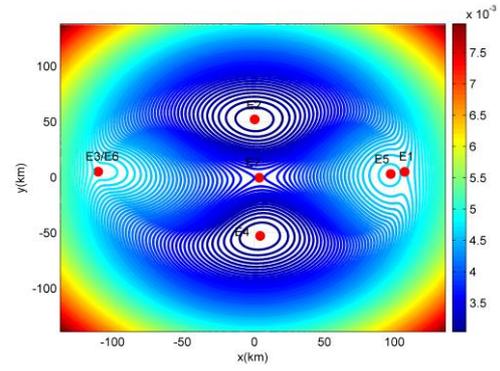

(b)

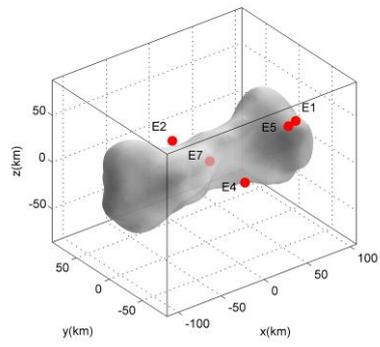 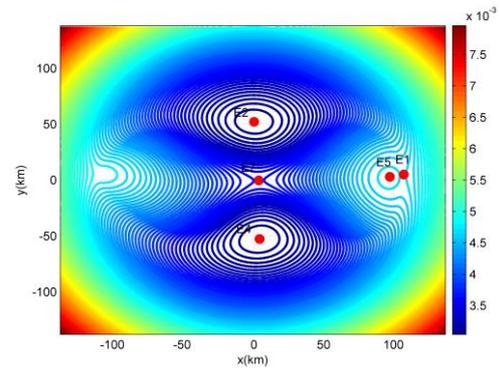

(c)

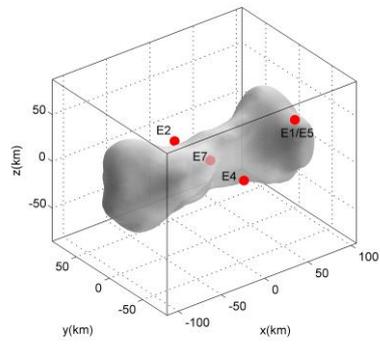 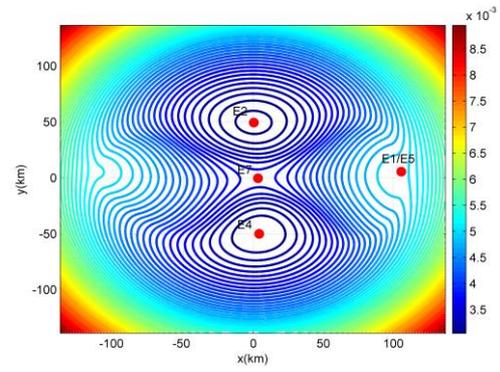

(d)

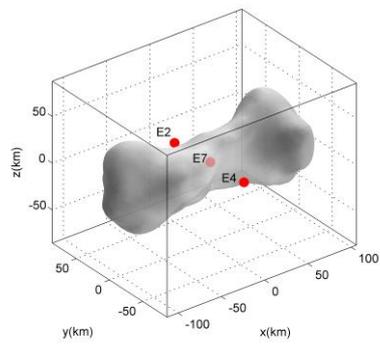 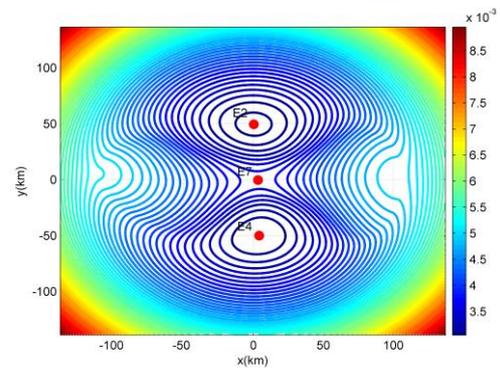

(e)

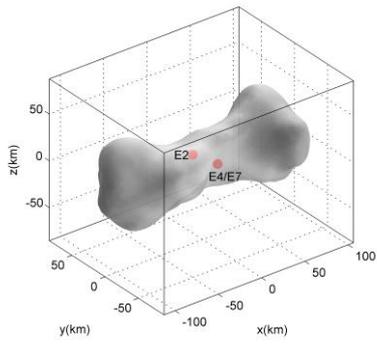
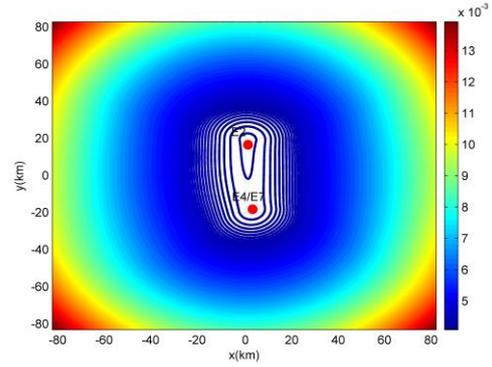

(f)

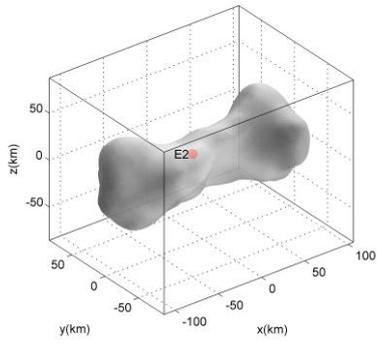
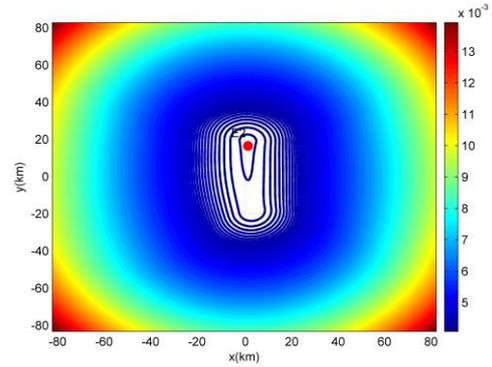

(g)

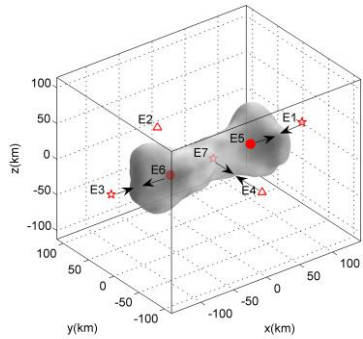

(h)

Figure 13 The annihilation of five equilibrium points of asteroid 216 Kleopatra as the rotational speed varies: (a) $\omega=1.0\omega_0$; (b) $\omega=1.944586\omega_0$; (c) $\omega=1.944587\omega_0$; (d) $\omega=2.03694\omega_0$; (e) $\omega=2.03695\omega_0$; (f) $\omega=4.270\,772\omega_0$; (g) $\omega=4.270\,773\omega_0$; (h) the variation trend of the equilibrium points.

### 3.2.2　2063 Bacchus

The annihilation classification of asteroid 2063 Bacchus is Type I for the seven equilibrium points of Kleopatra-shaped objects. Figure 14 shows the creation and annihilation of five equilibrium points of 2063 Bacchus as the rotational speed varies. At the beginning, $E6$ and $E7$ are created inside the body of the asteroid 2063 Bacchus when $\omega = 3.943\omega_0$. $E6$ belongs to Case O2 while $E7$ belongs to Case O1. During the creation, the Saddle-Node bifurcation occurs. After the creation of $E6$ and $E7$ there are three equilibrium points inside the body, $E5$, $E6$, and $E7$.

When the rotational speed is between $\omega = 3.943\omega_0$ and $\omega = 4.202\omega_0$, only three equilibrium points are inside the body, i.e., $E5$, $E6$, and $E7$. When $\omega = 4.202\omega_0$, $E1$ and $E6$ collide and annihilate each other. During the first annihilation the Saddle-Node bifurcation occurs. After the first annihilation there are five equilibrium points left, which are $E2$, $E3$, $E4$, $E5$, and $E7$, respectively. $E5$ and $E7$ are inside the body, and the others are outside.

As the rotational speed continues to vary, $E3$ and $E5$ move toward each other, they collide on the surface of the body and annihilate each other when $\omega = 4.527\omega_0$. During the second annihilation the Saddle-Node bifurcation occurs. After the second annihilation there are only three equilibrium points left, which are $E2$, $E4$, and $E7$, respectively. Only $E7$ is inside the body, $E2$ and $E4$ are outside the body. The topological cases of the equilibrium points do not change before the creation, the first annihilation, and the second annihilation.

The inside equilibrium point $E7$ and the outside equilibrium point $E4$ approach each other during the interval of $\omega = 4.527\omega_0$ to $\omega = 7.18\omega_0$. The topological case of equilibrium point $E2$ changes from Case O5 to Case O1 before the third annihilation. $E4$ and $E7$ collide on the surface of the body and annihilate each other when $\omega = 7.18\omega_0$. During the third annihilation the Saddle-Saddle bifurcation occurs. After the third annihilation only equilibrium point $E2$ is left on the surface of the body. As the rotational speed continues to vary, $E2$ moves into the body.

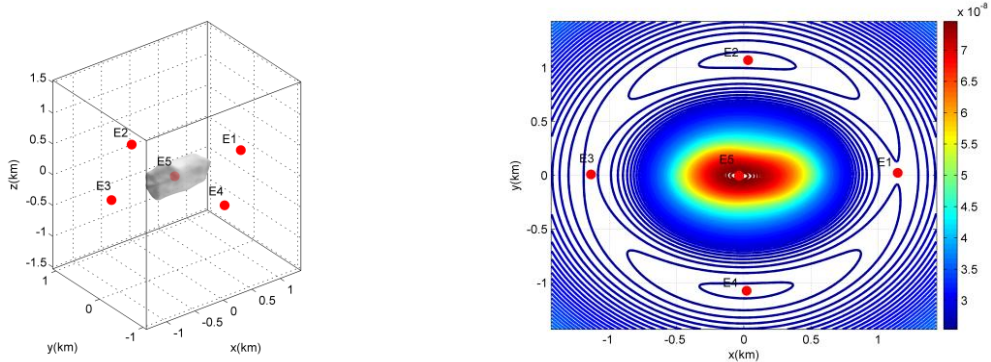

(a)

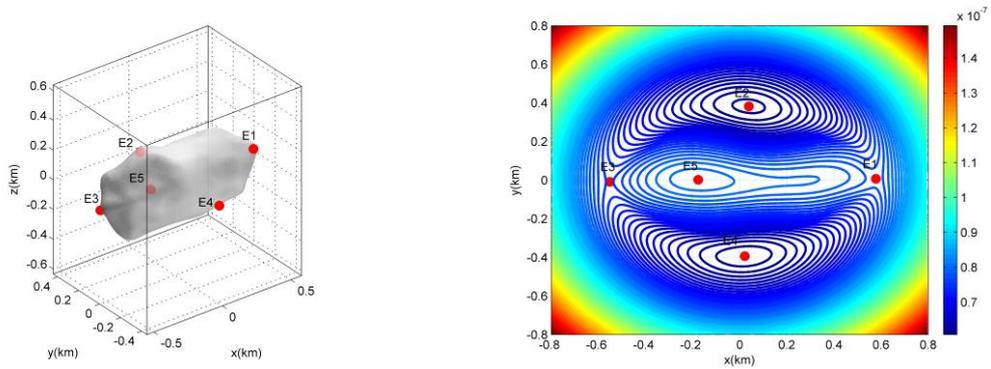

(b)

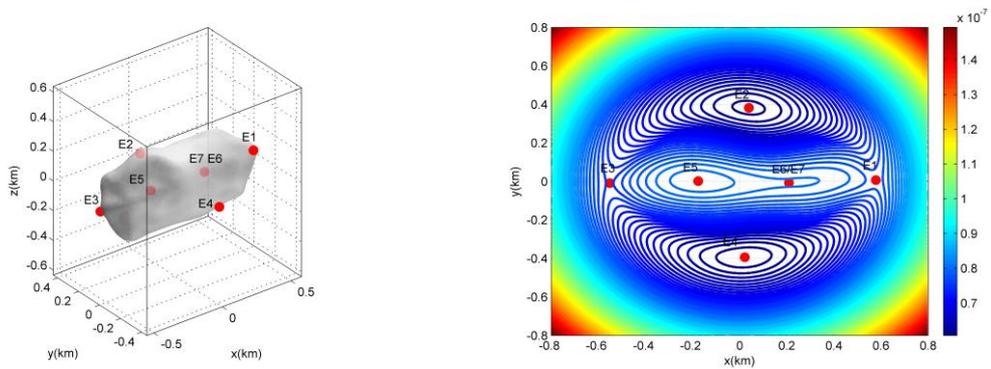

(c)

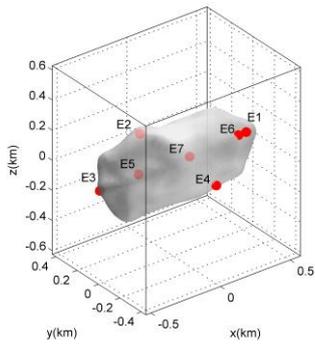 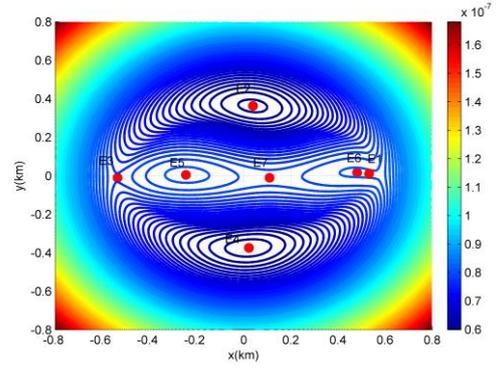

(d)

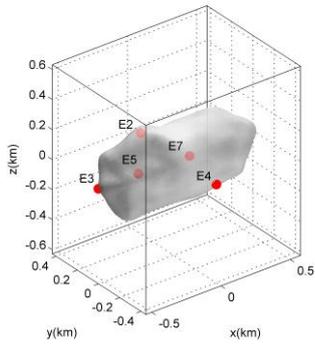 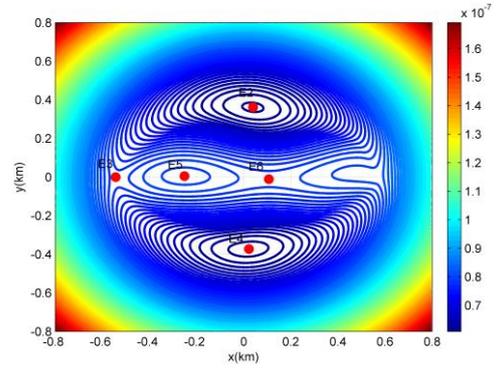

(e)

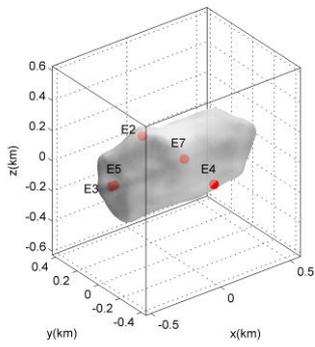 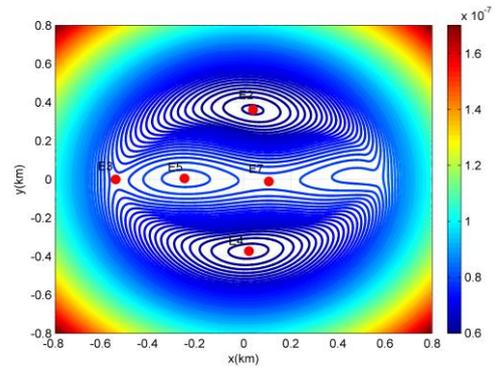

(f)

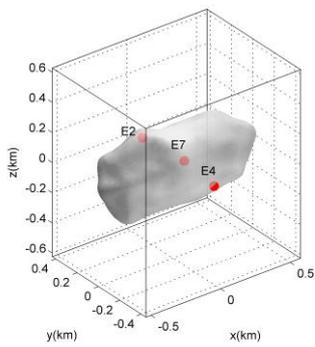 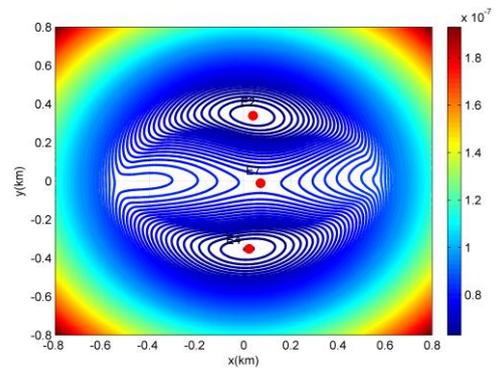

(g)

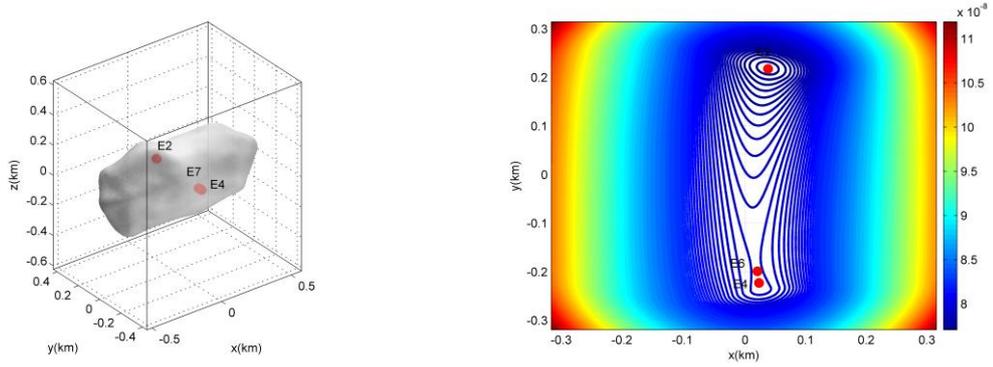

(h)

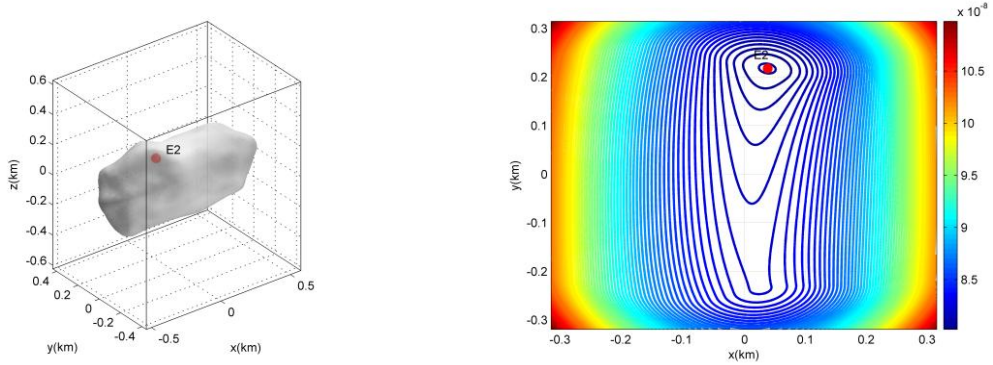

(i)

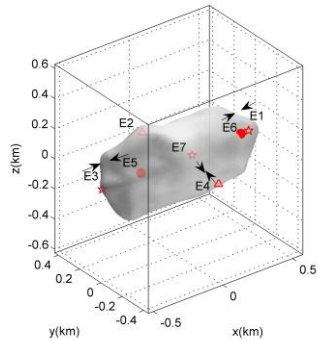

(j)

Figure 14 The annihilation of five equilibrium points of asteroid 2063 Bacchus as the rotational speed varies: (a) $\omega=1.0\omega_0$; (b) $\omega=3.943\omega_0$; (c) $\omega=3.944\omega_0$; (d) $\omega=4.202\omega_0$; (e) $\omega=4.203\omega_0$; (f) $\omega=4.527\omega_0$; (g) $\omega=4.528\omega_0$; (h) $\omega=7.15\omega_0$; (i) $\omega=7.19\omega_0$; (j) the variation trend of the equilibrium points.

### 3.2.3 101955 Bennu

The annihilation classification of asteroid 101955 Bennu is Type I for the nine equilibrium points. Figure 15 shows the creation and annihilation of five equilibrium points of 101955 Bennu as the rotational speed varies. During the first annihilation $E7$ and $E8$ approach each other and collide outside the asteroid 101955 Bennu. These two equilibrium points belong to Case O2 and Case O5, respectively. They annihilate when $\omega = 1.4124\omega_0$.

As the rotational speed varies from $\omega = 1.4124\omega_0$ to $\omega = 1.4178\omega_0$, $E1$ and $E9$ approach each other and collide when $\omega = 1.4178\omega_0$. The annihilation position is also outside the asteroid. After the annihilation there are only five equilibrium points left, which are $E2$, $E3$, $E4$, $E5$, and $E6$.

As the rotational speed continues to vary from $\omega = 1.4178\omega_0$ to $\omega = 1.4191\omega_0$, $E3$ and $E4$ approach each other and collide when $\omega = 1.4191\omega_0$. When the rotational speed gets larger than $\omega = 1.4191\omega_0$, $E3$ and $E4$ disappear. The topological cases of equilibrium points do not change before the first, second, and third annihilations.

The fourth annihilation occurs when $\omega = 1.4464\omega_0$; $E2$ and $E5$ collide and annihilate each other. The topological case of equilibrium point $E6$ changes from Case O5 to Case O1 before the fourth annihilation. After the annihilation only $E6$ is left.

Among the above four annihilations, only the second annihilation belongs to the Saddle-Node bifurcation, the others all belong to Saddle-Saddle bifurcation.

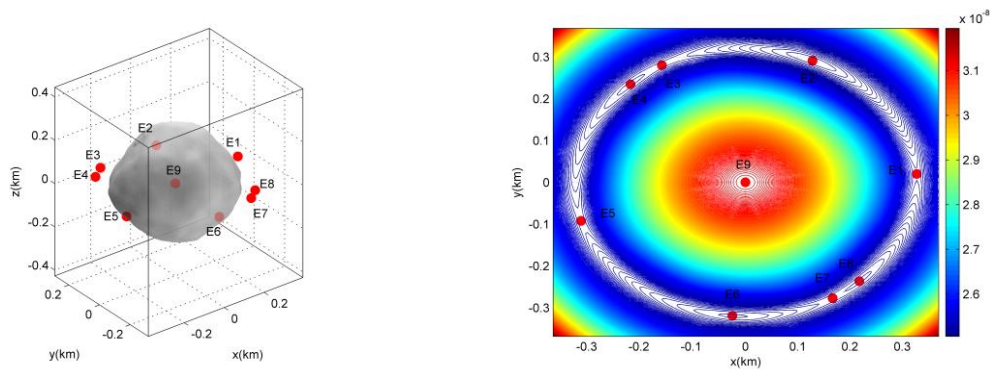

(a)

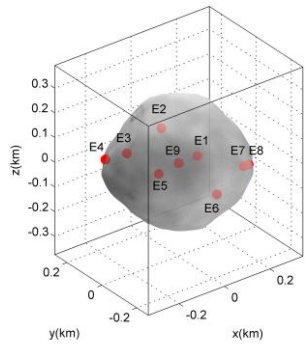 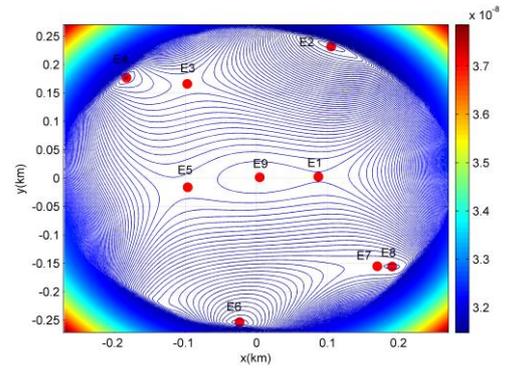

(b)

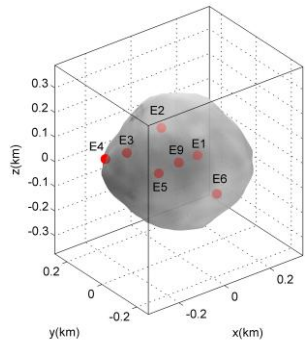 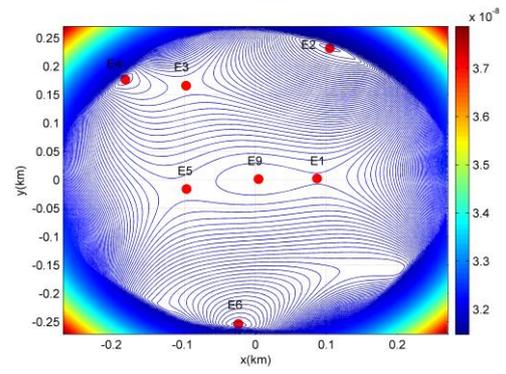

(c)

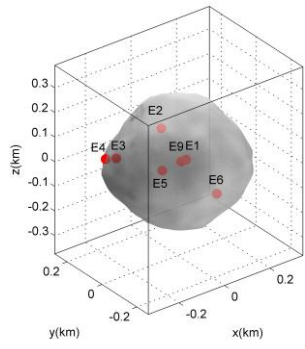 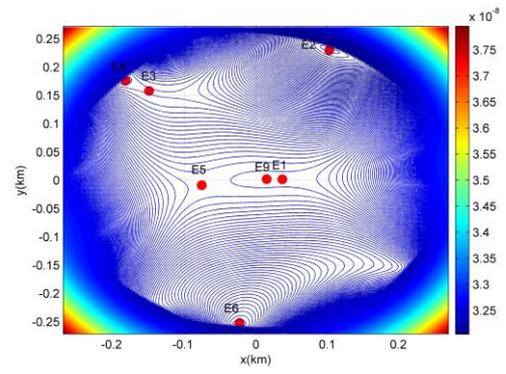

(d)

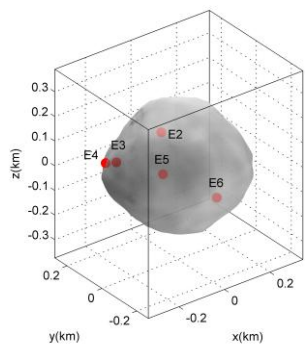 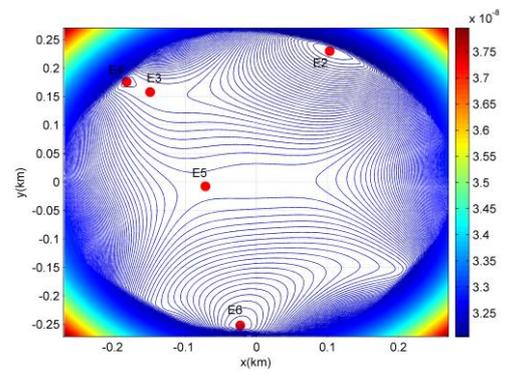

(e)

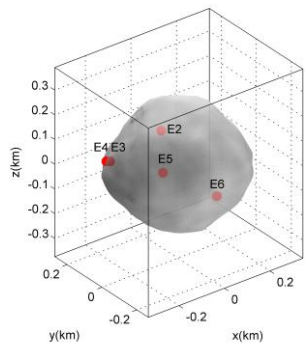 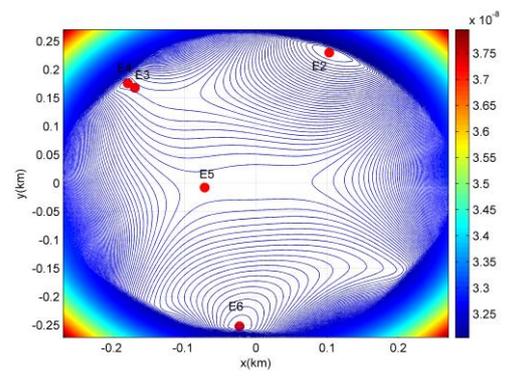

(f)

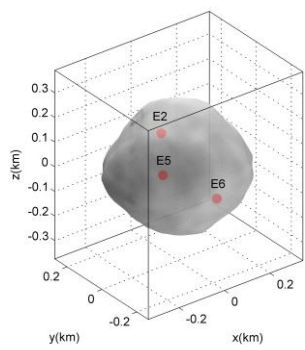 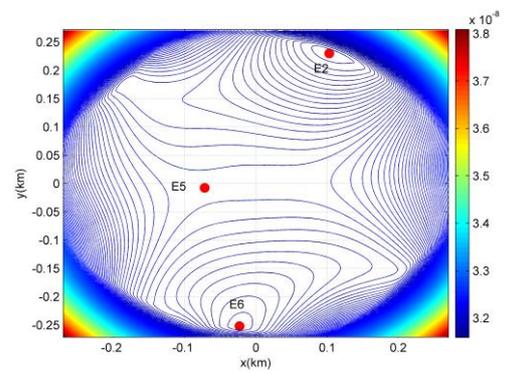

(g)

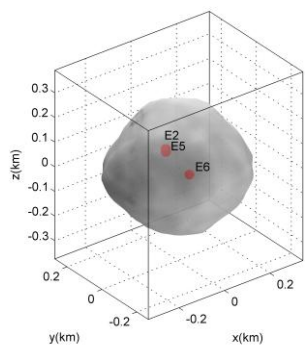 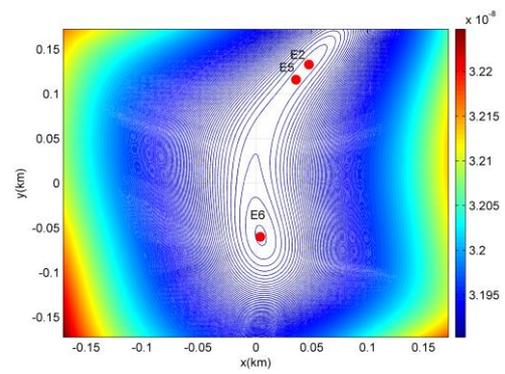

(h)

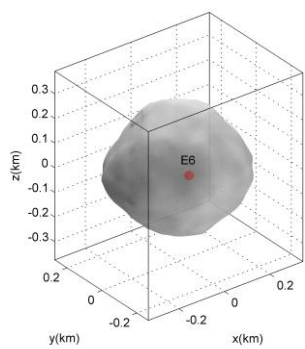 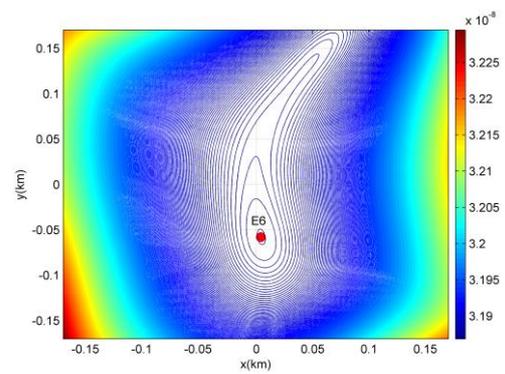

(i)

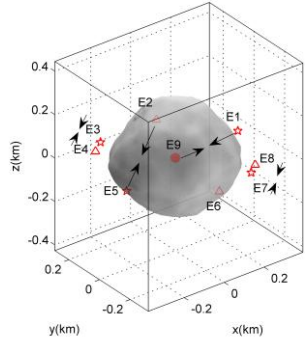

(j)

Figure 15 The annihilation of five equilibrium points of asteroid 101955 Bennu as the rotational speed varies: (a) $\omega=1.0\omega_0$; (b) $\omega=1.4124\omega_0$; (c) $\omega=1.4125\omega_0$; (d) $\omega=1.4178\omega_0$; (e) $\omega=1.4179\omega_0$; (f) $\omega=1.4191\omega_0$; (g) $\omega=1.4192\omega_0$; (h) $\omega=1.4464\omega_0$; (i) $\omega=1.4465\omega_0$ (j) the variation trend of the equilibrium points.

### 3.2.4  1682 Q1 Halley

The annihilation classification of comet 1682 Q1 Halley is Type I for the seven equilibrium points of Kleopatra-shaped objects. Figure 16 shows the creation and annihilation of the five equilibrium points of 1682 Q1 Halley as the rotational speed varies. $E6$ and $E7$ are created inside the body of the comet 1682 Q1 Halley when $\omega=7.64\omega_0$. $E6$ belongs to Case O2 while $E7$ belongs to Case O1. During creation, the Saddle-Node bifurcation occurs. After the creation of $E6$ and $E7$ three of the seven equilibrium points are inside the body, $E5$, $E6$, and $E7$ respectively.

When the rotational speed changes from $7.64\omega_0$ to $\omega=7.91\omega_0$, $E1$ and $E6$ approach each other. These two equilibrium points collide and annihilate each other on the surface of the comet 1682 Q1 Halley when $\omega=7.91\omega_0$. During the first annihilation the Saddle-Node bifurcation occurs. After the first annihilation five equilibrium points remain, i.e. $E2$, $E3$, $E4$, $E5$, and $E7$. Only $E5$ and $E7$ are inside the comet.

When the rotational speed changes from $\omega=7.91\omega_0$ to $\omega=9.36\omega_0$, $E3$ and $E5$ approach each other. These two equilibrium points collide and annihilate each other on the surface of the comet 1682 Q1 Halley when $\omega=9.36\omega_0$. During the second annihilation the Saddle-Node bifurcation occurs. After the second annihilation there are three equilibrium points left, i.e. $E2$, $E4$, and $E7$. Only $E7$ is inside the comet. The topological cases of the equilibrium points do not change before the creation, the first annihilation, and the second annihilation.

As the rotational speed continues to vary, $E2$ moves into the body and collides with $E7$ when $\omega=12.7\omega_0$. The topological case of equilibrium point $E4$ changes from Case O5 to Case O1 before the second annihilation. During the third annihilation the Saddle-Saddle bifurcation occurs. After the third annihilation only equilibrium point $E4$ is left.

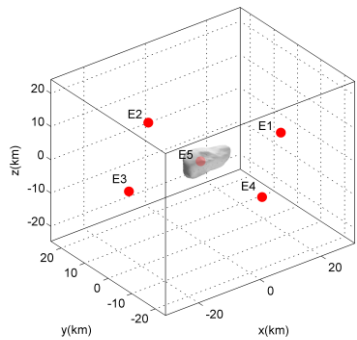 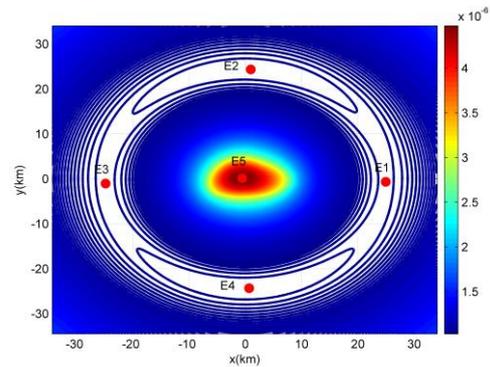

(a)

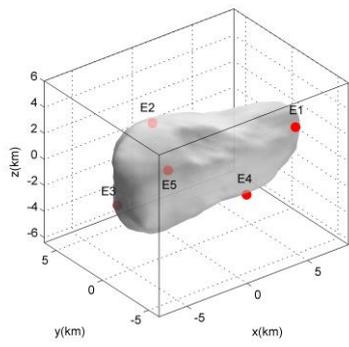
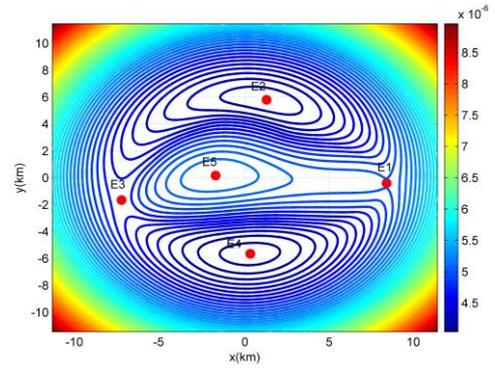

(b)

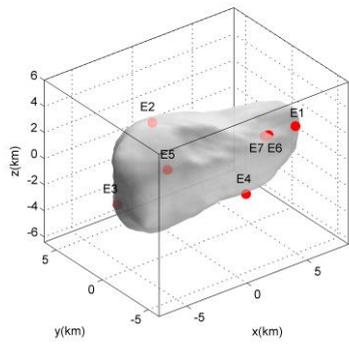
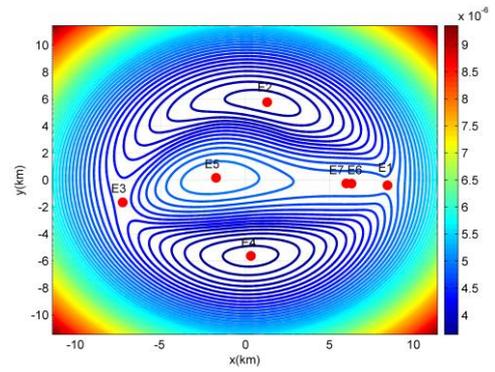

(c)

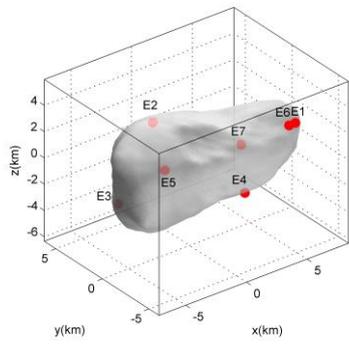
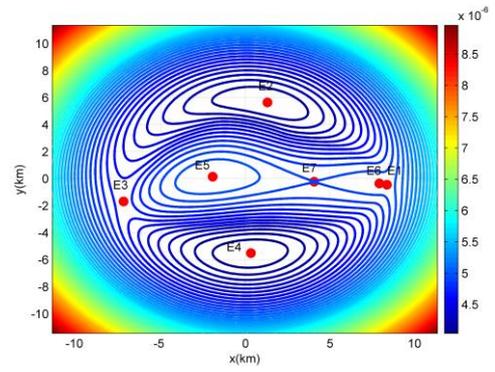

(d)

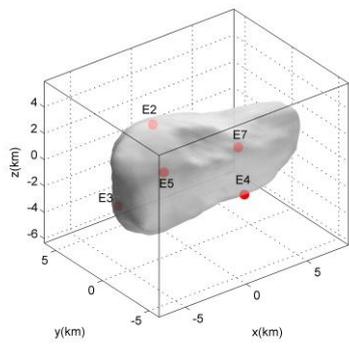
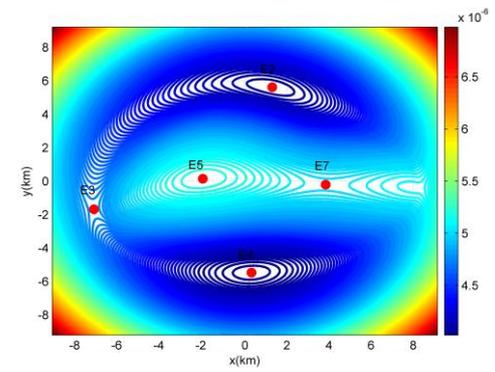

(e)

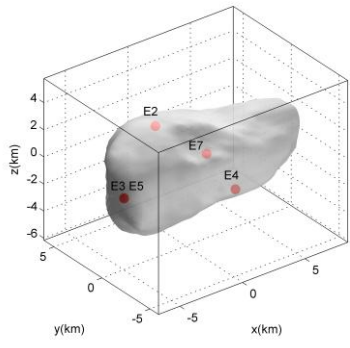 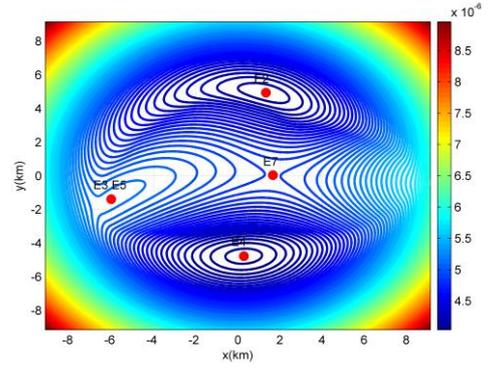

(f)

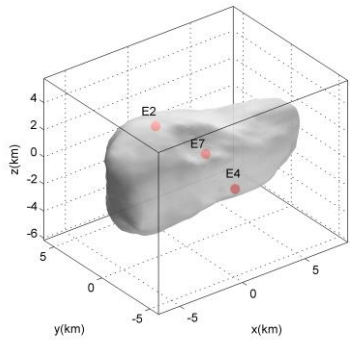 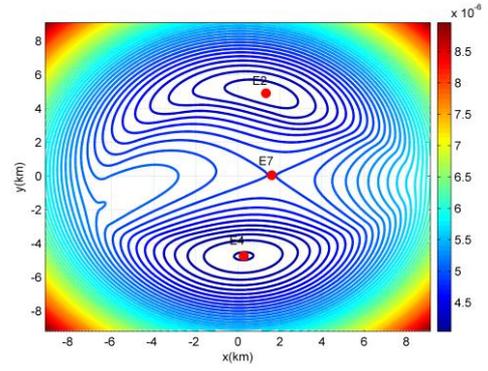

(g)

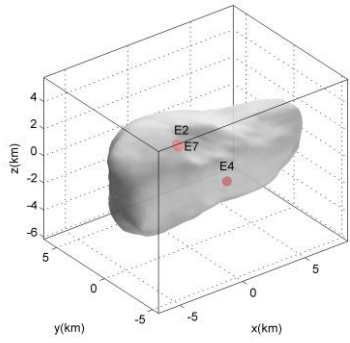 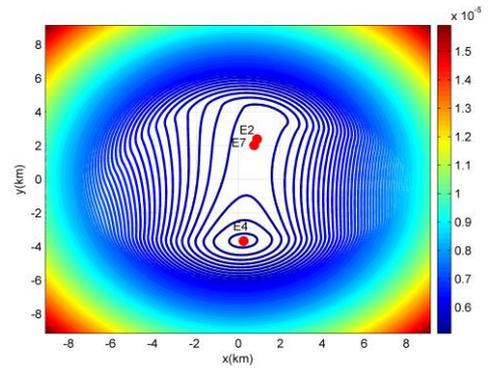

(h)

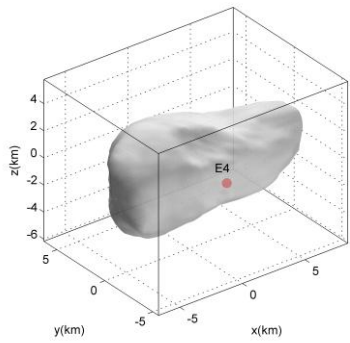 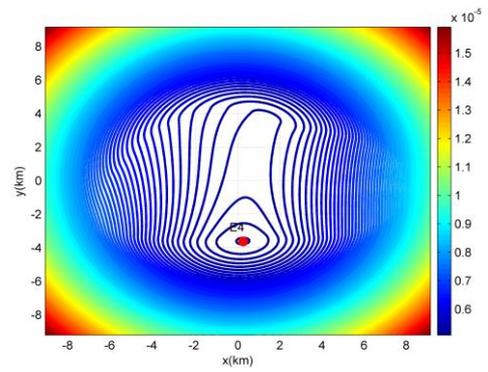

(i)

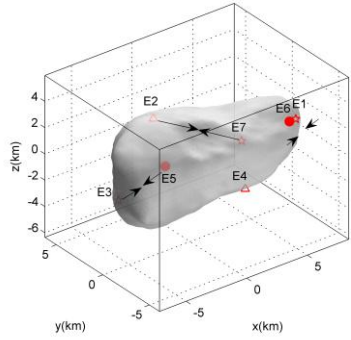

(j)

Figure 16 The annihilation of five equilibrium points of comet 1682 Q1/Halley as the rotational speed varies: (a) $\omega=1.0\omega_0$; (b) $\omega=7.64\omega_0$; (c) $\omega=7.65\omega_0$; (d) $\omega=7.91\omega_0$; (e) $\omega=7.92\omega_0$; (f) $\omega=9.36\omega_0$; (g) $\omega=9.27\omega_0$; (h) $\omega=12.7\omega_0$; (i) $\omega=12.8\omega_0$; (j) the variation trend of the equilibrium points.

## 3.3 Annihilate with Surface Equilibria

The smooth surface equilibria satisfies

$$\nabla V(\mathbf{r}_s) \cdot \nabla W(\mathbf{r}_s) = 0. \tag{3}$$

Where $\mathbf{r}_s$ represents the position vector from the mass centre of the minor celestial body to a point on the surface, $W(\mathbf{r}_s)$ represents the surface shape function of the body. Using Eq. (3), one can calculate the positions of the smooth surface equilibria. From Table A2, one can see that for several minor celestial bodies, the annihilation positions may on the surface of the bodies.

The equilibrium point in the gravitational field of the body moves during the variety of the rotational speed. When the equilibrium point touches the surface of the body, the equilibrium point is also the smooth surface equilibrium, i.e. the equilibrium point and the smooth surface equilibrium collides. For the annihilation position of the equilibrium points in the gravitational field of the body which is on the surface, at least three points collide at the annihilation position, the first one is the equilibrium point from the external of the body, the second one is the equilibrium point from the internal of the body, and the third one is the smooth surface equilibrium.

We now take asteroid 216 Kleopatra for instance, for other minor celestial bodies, the smooth surface equilibrium are also exist and the equilibrium point and the smooth surface equilibrium also collides during the variety of the rotational speed. From the above Table A2, we know that all the three annihilation positions of asteroid 216 Kleopatra during the variety of the rotational speed are on the surface of the body. We now calculate the smooth surface equilibria of 216 Kleopatra. Because the shape function of the asteroid is generated by the polyhedron method, the surface equilibria can not strictly satisfies Eq. (3). The relative error is set to be $0.8 \times 10^{-4}$, i.e., if the position $\mathbf{r}_s$ satisfies $\nabla V(\mathbf{r}_s) \cdot \nabla W(\mathbf{r}_s) \leq 0.8 \times 10^{-4}$, the position is a surface equilibrium. Figure 17 presents contour plot of surface effective potential and relative positions of smooth surface equilibria and shape of asteroid 216 Kleopatra. To show the relative value, in Figure 17(a), the colour bar 1.0 means the maximum value while the colour bar 0.0 means the minimum value. The length unit and time unit are set to be 219.0361 km and 5.385 h, respectively. The maximum value and the minimum value of the surface effective potential are -324.794 and -3967.60. Using the relative error $0.8 \times 10^{-4}$, the surface equilibria are calculated to be regions on the surface of the asteroid 216 Kleopatra. When the equilibrium point touches the surface of the body during the variety of the gravitational parameters, the equilibrium point can only touch the surface equilibria are regions. From Figure 13 and Table A2, one can see that the equilibrium points E3 and E6 touch the surface equilibria in the –x half-space, which can be seen in Figure 17(d). The equilibrium points E1 and E5 touch the surface equilibria in the +x half-space, which can be seen in Figure 17(c). The equilibrium points E4 and E7 touch the surface equilibria in the -y half-space, which can be seen in Figure 17(f).

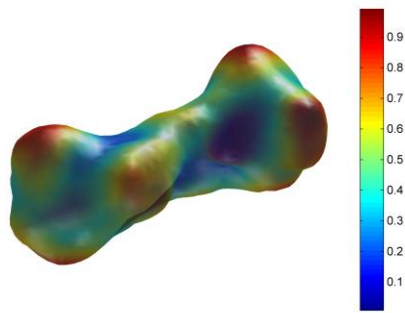

(a)

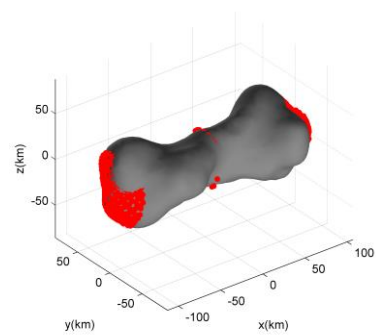

(b)

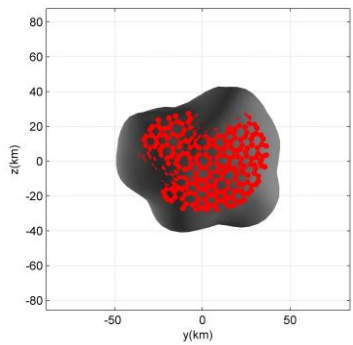

(c)

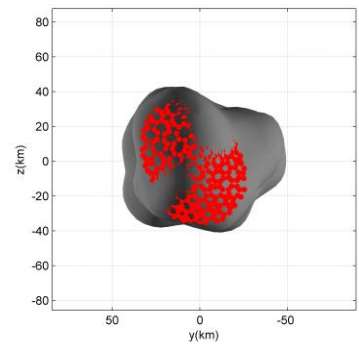

(d)

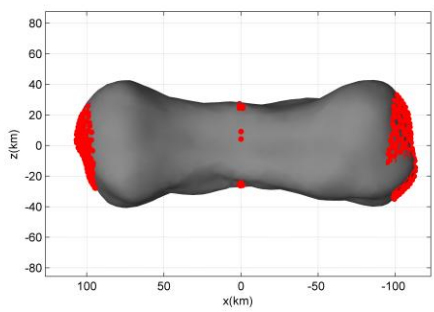

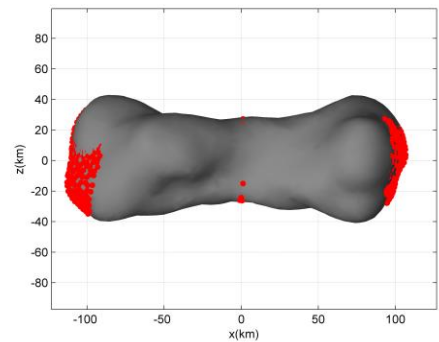

(e) 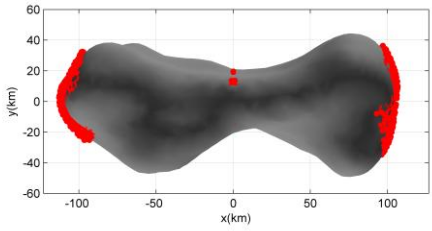 (f) 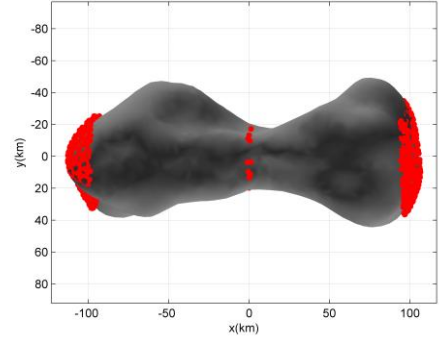

(g) (h)

Figure 17 Surface effective potential and smooth surface equilibria of asteroid 216 Kleopatra: (a) Surface effective potential; (b) 3D view of smooth surface equilibria; (c) viewed from +x axis; (d) viewed from -x axis; (e) viewed from +y axis; (f) viewed from -y axis; (g) viewed from +z axis; (h) viewed from -z axis.

## 4. Conclusions

The annihilation classifications of relative equilibrium points in the gravitational potential of irregular-shaped minor celestial bodies were investigated. The topological classifications and indices of equilibrium points were discussed. The equilibrium points with different indices have a staggered distribution. If the parameters (such as rotational speed) vary, equilibria may annihilate each other. The annihilation classifications not only depend on the parameters but also on the number and topological cases of equilibria.

There are seven types of annihilation classifications for equilibria of minor bodies when they have five equilibrium points. We also discussed the types of annihilation classifications for the seven equilibria of Kleopatra-shaped objects and the nine equilibria of Bennu-shaped objects. We calculated the annihilations and creations of relative equilibria in the gravitational field of ten minor bodies, including eight asteroids, one satellite of a planet, and one cometary nucleus. Only two objects have both annihilations and creations when the rotational speed increases, the asteroid 2063 Bacchus, and the comet Q1/Halley. Others have only annihilations as the rotational speed increases. The Saddle-Node bifurcations occur when the equilibrium points created for asteroid 2063 Bacchus and comet Q1/Halley. After the creation of the equilibrium points, both asteroid 2063 Bacchus and comet Q1/Halley have seven equilibrium points; the number of equilibrium points is the same as that for 216 Kleopatra. The first and second annihilations of asteroid 216 Kleopatra, 2063 Bacchus, and comet Q1/Halley are Saddle-Node bifurcations, while the third annihilation is a Saddle-Saddle bifurcation. The annihilation classification of asteroid 216 Kleopatra and 2063 Bacchus shows that they are Type I-a for the seven equilibrium points of Kleopatra-shaped objects. The annihilation classification of comet Q1/Halley is Type I for the seven equilibrium points of Kleopatra-shaped objects.

In addition, we find that all of the annihilations are collisional annihilations. As the rotational speed increases, each of the asteroids 243 Ida, 951 Gaspra, and 1620 Geographos, as well as the satellite of planet S16 Prometheus have two annihilations; the first annihilation is a Saddle-Node bifurcation, while the second one is a Saddle-Saddle bifurcation. As the rotational speed increases, both of the asteroids 2867 Steins and 6489 Golevka have two annihilations, and both of the two annihilations are Saddle-Node bifurcations. The annihilation classification of 1620 Geographos, 951 Gaspra, 243 Ida, and S16 Prometheus are Type III for five equilibrium points. The annihilation classification of 2867 Steins is Type IV, while 6489 Golevka is Type I for five equilibrium points.

For the asteroid 101955 Bennu, which has nine equilibrium points, there are four annihilations during the increase in the rotational speed; only the second annihilation is a Saddle-Node bifurcation; the others are all Saddle-Saddle bifurcations. The annihilation classification of 101955 Bennu is also presented.

It seems that for some minor bodies in some cases, there is a link between an equilibrium point that touches the surface and the collisions between equilibria. This link includes the three annihilations of equilibria of 216 Kleopatra, the two annihilations of equilibria of 951 Gaspra, and the second annihilations of equilibria of 243 Ida. However, there are also several cases when the equilibrium point touching the surface and the collisions between equilibria are not simultaneous; for instance, the two annihilations of equilibria of 2867 Steins. If the annihilation position is on the surface of the body, there are at least three points collide at the annihilation position, including the equilibrium point from the external of the body, the equilibrium point from the internal of the body, and the smooth surface equilibrium.

Further research will be performed to provide a way to estimate when these annihilations and generations occur or how to detect them. We will also investigate whether even slight internal perturbations in the minor celestial bodies may contribute to a noticeable variety of equilibrium points. The internal perturbations means the variety of the mascon and the gap.

**Acknowledgments**

This research was supported by the National Natural Science Foundation of China (No. 11772356)，the National Science Foundation for Distinguished Young Scholars (11525208), and China Postdoctoral Science Foundation- General Program (No. 2017M610875).

**Appendix A**

Table A1 Physical properties of irregular minor celestial bodies

| Serial number | Minor bodies | Diameters (km) | Bulk density ($g \cdot cm^{-3}$) | Rotation period (h) |
|---|---|---|---|---|
| 1 | 216 Kleopatra[a1,a2] | 217×94×81 | 3.6 | 5.385 |
| 2 | 243 Ida[b1,b2] | 59.8×25.4×18.6 | 2.6 | 4.63 |
| 3 | 951 Gaspra[c1,c2,c3] | 18.2×10.5×8.9 | 2.71 | 7.042 |
| 4 | 1620 Geographos[d] | 5.39×2.40×2.02 | 2.0 | 5.223 |
| 5 | 2063 Bacchus[e] | 1.11×0.53×0.50 | 2.0 | 14.9 |
| 6 | 2867 Steins[f1,f2] | 6.67×5.81×4.47 | 1.8 | 6.04679 |
| 10 | 6489 Golevka[g1,g2] | 0.75×0.55×0.59 | 2.7 | 6.026 |
| 8 | 101955 Bennu[h] | 0.58×0.44×0.53 | 0.97 | 4.288 |
| 9 | S16 Prometheus[i1,i2] | 148×93×72 | 0.48 | 14.71 |
| 10 | 1P/Halley[j1,j2] | 16.8×8.77×7.76 | 0.6 | 52.8 |

[a1] Carry et al. 2012. [a2] Ostro et al. 2000. [b1] Wilson et al. 1999. [b2] Vokrouhlický et al. 2003. [c1] Krasinsky et al. 2002. [c2] Kaasalainen et al. 2001. [c3] Stooke 1995. [d] Ostro et al. 1996. [e] Benner et al. 1999. [f1] Jorda et al. 2012. [f2] Keller et al. 2010. [g1] Chesley et al. 2003. [g2] Hudson et al. 2000. [h] Nolan et al. 2007. [i1] Thomas et al. 2010. [i2] Spitale et al. 2006. [j1] Sagdeev et al. 1988. [j2] Peale and Lissauer 1989.

Table A2 Positions of Annihilation of Equilibria around Minor Celestial Bodies
NEA: Number of Equilibria after Annihilations, PA: Positions of Annihilation

**216 Kleopatra**

| Index | Annihilation/Generation | Angular Speed | Bifurcations | NEA | PA |
|---|---|---|---|---|---|
| 1 | E3/E6→ Annihilation | $\omega=1.944586\omega_0$ | Saddle-Node | 5 | Surface |

| 2 | E1/E5→ Annihilation | $\omega=2.03694\omega_0$ | Saddle-Node | 3 | Surface |
|---|---|---|---|---|---|
| 3 | E4/E7→ Annihilation | $\omega=4.270772\omega_0$ | Saddle-Saddle | 1 | Surface |

### 243 Ida

| Index | Annihilation/Generation | Angular Speed | Bifurcations | NEA | PA |
|---|---|---|---|---|---|
| 1 | E1/E5→ Annihilation | $\omega=1.4992\omega_0$ | Saddle-Node | 3 | Inside |
| 2 | E3/E4→ Annihilation | $\omega=2.24719\omega_0$ | Saddle-Saddle | 1 | Surface |

### 951 Gaspra

| Index | Annihilation/Generation | Angular Speed | Bifurcations | NEA | PA |
|---|---|---|---|---|---|
| 1 | E3/E5→ Annihilation | $\omega=2.9586\omega_0$ | Saddle-Node | 3 | Surface |
| 2 | E1/E4→ Annihilation | $\omega=3.8565\omega_0$ | Saddle-Saddle | 1 | Surface |

### 1620 Geographos

| Index | Annihilation/Generation | Angular Speed | Bifurcations | NEA | PA |
|---|---|---|---|---|---|
| 1 | E1/E5→ Annihilation | $\omega=1.4745\omega_0$ | Saddle-Node | 3 | Inside |
| 2 | E2/E3→ Annihilation | $\omega=2.365\omega_0$ | Saddle-Saddle | 1 | Surface |

### 2063 Bacchus

| Index | Annihilation/Generation | Angular Speed | Bifurcations | NEA | PA |
|---|---|---|---|---|---|
| 1 | Generation→E6/E7 | $\omega=3.943\omega_0$ | Saddle-Node | 7 | Inside |
| 2 | E1/E6→ Annihilation | $\omega=4.202\omega_0$ | Saddle-Node | 5 | Inside |
| 3 | E3/E5→ Annihilation | $\omega=4.527\omega_0$ | Saddle-Node | 3 | Inside |
| 4 | E4/E7→ Annihilation | $\omega=7.18\omega_0$ | Saddle-Saddle | 1 | Surface |

### 2867 Steins

| Index | Annihilation/Generation | Angular Speed | Bifurcations | NEA | PA |
|---|---|---|---|---|---|
| 1 | E1/E5→ Annihilation | $\omega=2.972\omega_0$ | Saddle-Node | 3 | Inside |
| 2 | E2/E3→ Annihilation | $\omega=3.26\omega_0$ | Saddle-Node | 1 | Inside |

### 6489 Golevka

| Index | Annihilation/Generation | Angular Speed | Bifurcations | NEA | PA |
|---|---|---|---|---|---|
| 1 | E1/E5→ Annihilation | $\omega=2.696\omega_0$ | Saddle-Node | 3 | Inside |
| 2 | E2/E3→ Annihilation | $\omega=3.437\omega_0$ | Saddle-Node | 1 | Surface |

### 101955 Bennu

| Index | Annihilation/Generation | Angular Speed | Bifurcations | NEA | PA |
|---|---|---|---|---|---|

| 1 | E7/E8→Annihilation | $\omega=1.4124\omega_0$ | Saddle-Saddle | 7 | Inside |
| 2 | E1/E9→Annihilation | $\omega=1.4178\omega_0$ | Saddle-Node | 5 | Inside |
| 3 | E3/E4→Annihilation | $\omega=1.4191\omega_0$ | Saddle-Saddle | 3 | Surface |
| 4 | E2/E5→Annihilation | $\omega=1.4464\omega_0$ | Saddle-Saddle | 1 | Outside |

**S16 Prometheus**

| Index | Annihilation/Generation | Angular Speed | Bifurcations | NEA | PA |
|---|---|---|---|---|---|
| 1 | E3/E5→Annihilation | $\omega=2.270\omega_0$ | Saddle-Node | 3 | Inside |
| 2 | E1/E2→Annihilation | $\omega=3.068\omega_0$ | Saddle-Saddle | 1 | Inside |

**1682 Q1 Halley**

| Index | Annihilation/Generation | Angular Speed | Bifurcations | NEA | PA |
|---|---|---|---|---|---|
| 1 | Generation→E6/E7 | $\omega=7.64\omega_0$ | Saddle-Node | 7 | Inside |
| 2 | E1/E6→Annihilation | $\omega=7.91\omega_0$ | Saddle-Node | 5 | Surface |
| 3 | E3/E5→Annihilation | $\omega=9.36\omega_0$ | Saddle-Node | 3 | Surface |
| 4 | E2/E7→Annihilation | $\omega=12.7\omega_0$ | Saddle-Saddle | 1 | Inside |

**References**

Benner, L. A. M., Hudson, R. S., Ostro, S. J., Rosema, K. D., Giorgini, J. D., Yeomans, D. K., Jurgens, R. F., Mitchell, D. L., Winkler, R., Rose, R., Slade, M. A., Thomas, M. L., Pravec, P., 1999. Radar observations of asteroid 2063 Bacchus. Icarus 139(2), 309-327.

Carry, B., 2012. Density of asteroids. Planet. Space Sci. 73(1), 98-118.

Chanut, T. G. G., Winter, O. C., Tsuchida, M., 2014. 3D stability orbits close to 433 Eros using an effective polyhedral model method. Mon. Not. R. Astron. Soc. 438(3), 2672-2682.

Chanut, T. G. G., Winter, O. C., Amarante, A., Araújo, N. C. S., 2015. 3D plausible orbital stability close to asteroid (216) Kleopatra. Mon. Not. R. Astron. Soc. 452(2). 1316-1327.

Chesley, S. R., Ostro, S. J., Vokrouhlicky, D., Capek, D., Giorgini, J. D., Nolan, M. C., Margot, J. L., Hine, A. A., Benner, L. A., Chamberlin, A. B., 2003. Direct detection of the Yarkovsky effect by radar ranging to asteroid 6489 Golevka. Science 302(5651), 1739-1742.

Cotto-Figueroa, D., Statler, T. S., Richardson, D. C., Tanga, P., 2014. Coupled spin and shape evolution of small rubble-pile asteroids: self-limitation of the YORP effect. Astrophys. J. 803(1): 25.

Guirao, J. L. G., Rubio, R. G., Vera, J. A., 2011. Nonlinear stability of the equilibria in a double-bar rotating system. J. Comput. Appl. Math. 235(7), 1819-1825.

Hirabayashi, M., 2015. Failure modes and conditions of a cohesive, spherical body due to YORP spin-up. Mon. Not. R. Astron. Soc. 454, 2249-2257.

Hirabayashi, M., Sánchez, D. P., Scheeres. D. J., 2015. Failure modes and conditions of a cohesive, spherical body due to YORP spin-up. Mon. Not. R. Astron. Soc. 454(2), 2249-2257.

Hirabayashi, M., Scheeres, D. J., 2014. Analysis of asteroid (216) Kleopatra using dynamical and structural constraints. Astrophys. J. 780(2), 386-406.

Holsapple, K. A., 2004. Equilibrium figures of spinning bodies with self-gravity. Icarus 172, 272-303.

Hudson, R. S., Ostro, S. J., Jurgens, R. F., Rosema, K. D., Giorgini, J. D., Winkler, R., Rose, R., Choate, D., Cormier, R. A., Franck, C. R., Frye, R., Howard, D., Kelley, D., Littlefair, R., Slade, M. A., Benner, L. A. M., Thomas, M. L., Mitchell, D. L., Chodas, P. W., Yeomans, D. K., Scheeres, D. J., Palmer, P., Zaitsev, A., Koyama, Y., Nakamura, A., Harris, A. W., Meshkov, M. N., 2000. Radar observations and physical model of asteroid 6489 Golevka. Icarus 148(1), 37-51.

Jiang, Y., Baoyin, H., Li, J., Li, H., 2014. Orbits and manifolds near the equilibrium points around a rotating asteroid. Astrophys. Space Sci. 349(1), 83-106.

Jiang, Y., Baoyin, H., Li, H., 2015. Collision and annihilation of relative equilibrium points around asteroids with a changing parameter. Mon. Not. R. Astron. Soc. 452 (4): 3924-3931.

Jiang, Y., Baoyin, H., Wang, X., Yu Y., Li, H., Peng C., Zhang Z., 2016a, Order and chaos near equilibrium points in the potential of rotating highly irregular-shaped celestial bodies. Nonlinear Dynam. 83(1), 231-252.

Jiang, Y., Zhang, Y., Baoyin, H., 2016b, Surface motion relative to the irregular celestial bodies. Planet. Space Sci. 127, 33-43.

Jorda, L., Lamy, P. L., Gaskell, R. W., Kaasalainen, M., Groussin, O., Besse, S., Faury, G., 2012. Asteroid (2867) Steins: Shape, topography and global physical properties from OSIRIS observations. Icarus 221, 1089-1100.

Kaasalainen, M., Torppa, J., Muinonen, K., 2001. Optimization methods for asteroid lightcurve inversion: II. The complete inverse problem. Icarus 153(1), 37-51.

Keller, H. U., Barbieri, C., Koschny, D., Lamy, P., Rickman, H., Rodrigo, R., Sierks, H., A'Hearn, M. F., Angrilli, F., Barucci, M. A., Bertaux, J. L., Cremonese, G., Da Deppo, V., Davidsson, B., De Cecco, M., Debei, S., Fornasier, S., Fulle, M., Groussin, O., Gutierrez, P. J., Hviid, S. F., Ip, W. H., Jorda, L., Knollenberg, J., Kramm, J. R., Kuhrt, E., Kuppers, M., Lara, L. M., Lazzarin, M., Moreno, J. L., Marzari, F., Michalik, H., Naletto, G., Sabau, L., Thomas, N., Wenzel, K. P., Bertini, I., Besse, S., Ferri, F., Kaasalainen, M., Lowry, S., Marchi, S., Mottola, S., Sabolo, W., Schroder, S. E., Spjuth, S., Vernazza, P., 2010. E-type Asteroid (2867) Steins as imaged by OSIRIS on board Rosetta. Science 327(5962), 190-193.

Krasinsky, G. A., Pitjeva, E. V., Vasilyev, M. V., Yagudina, E. I., 2002. Hidden Mass in the Asteroid Belt. Icarus 158(1), 98-105.

Lowry, S. C., Weissman, P. R., Duddy, S. R., Rozitis, B., Fitzsimmons, A., Green, S. F., Hicks, M. D., Snodgrass, C., Wolters, S. D., Chesley, S. R., Pittichová J., van Oers, P., 2014. The internal structure of asteroid (25143) Itokawa as revealed by detection of YORP spin-up. Astron. Astrophys. 562(2), 118-130.

Micheli, M., Paolicchi, P., 2008. YORP effect on real objects. I. statistical properties.

Astron. Astrophys. 490(1), 387-391.

Nolan, M. C., Magri, C., Ostro, S. J., Benner, L. A., Giorgini, J. D., Howell, E. S., Hudson, R. S., 2007. The Shape and Spin of 101955 (1999 RQ36) from Arecibo and Goldstone Radar Imaging. Bulletin of the American Astronomical Society 39, 433.

Ostro, S. J., Hudson, R. S., Nolan, M. C., Margot, J. L., Scheeres, D. J., Campbell, D. B., Magri, C., Giorgini, J. D., Yeomans, D. K., 2000. Radar observations of asteroid 216 Kleopatra. Science 288(5467), 836-839.

Ostro, S. J., Jurgens, R. F., Rosema, K. D., Hudson, R. S., Giorgini, J. D., Winkler, R., Yeomans, D. K., Choate, D., Rose, R., Slade, M. A., Howard, S. D., Scheeres, D. J., Mitchell, D. L., 1996. Radar observations of asteroid 1620 Geographos. Icarus 121(1), 46-66.

Palacián, J. F., Yanguas, P., Gutiérrez-Romero, S., 2006. Approximating the invariant sets of a finite straight segment near its collinear equilibria. SIAM J. Applied Dynam. Syst. 5(1), 12-29.

Peale, S. J., Lissauer, J. J., 1989. Rotation of Halley's comet. Icarus 79(2), 396-430.

Riaguas, A., Elipe, A., López-Moratalla, T., 2001. Non-linear stability of the equilibria in the gravity field of a finite straight segment. Celest. Mech. Dynam. Astron. 81(3), 235-248.

Richardson, D. C., Elankumaran, P., Sanderson, R. E., 2005. Numerical experiments with rubble piles: equilibrium shapes and spins. Icarus 173(2), 349-361.

Sagdeev, R. Z., Elyasberg, P. E., Moroz, V. I., 1988. Is the nucleus of comet Halley a low density body. Nature 331(6153), 240-242.

Scheeres, D. J., 2007. The dynamical evolution of uniformly rotating asteroids subject to YORP. Icarus 188(2), 430-450.

Scheeres, D. J., Hesar, S. G., Tardivel, S., Hirabayashi, M., Farnocchia, D., Mcmahon, J. W., 2016. The geophysical environment of Bennu. Icarus 276, 116-140.

Ševeček, P., Brož, M., Čapek, D., Ďurech. J., 2015. The thermal emission from boulders on (25143) Itokawa and general implications for the YORP effect. Mon. Not. R. Astron. Soc. 450(2), 2104-2115

Sharma, I., Jenkins, J. T., Burns, J. A., 2009. Dynamical passage to approximate equilibrium shapes for spinning, gravitating rubble asteroids. Icarus 200(1), 304-322.

Spitale, J. N., Jacobson, R. A., Porco, C. C., Owen, Jr. W. M., 2006. The Orbits of Saturn's Small Satellites Derived from Combined Historic and Cassini Imaging Observations. Astron. J. 132(2), 692-710.

Stooke, P. J., 1995. The surface of asteroid 951 Gaspra. Earth, Moon, and Planets, 75(1), 53-75.

Taylor, P. A., Jean-Luc, M., David, V., Scheeres, D. J., Petr, P., Lowry, S. C., et al., 2007. Spin rate of asteroid (54509) 2000 PH5 increasing due to the YORP effect. Science 316(5822), 274-277.

Thomas, P. C., 2010. Sizes, shapes, and derived properties of the Saturnian satellites after the Cassini nominal mission. Icarus 208(1), 395-401.

Vasilkova, O., 2005. Three-dimensional periodic motion in the vicinity of the equilibrium points of an asteroid. Astron. Astrophys. 430, 713-723.

Vokrouhlický, D., Čapek. D., 2002. YORP-induced long-term evolution of the spin state of small asteroids and meteoroids: rubincam's approximation. Icarus 159(2), 449-467.

Vokrouhlický, D., Nesvorný, D., Bottke, W. F., 2003. The vector alignments of asteroid spins by thermal torques. Nature 425(6954), 147-151.

Walsh, K. J., Richardson, D. C., Michel, P., 2012. Spin-up of rubble-pile asteroids: disruption, satellite formation, and equilibrium shapes. Icarus 220(2), 514-529.

Wang, X., Jiang, Y., Gong, S., 2014. Analysis of the potential field and equilibrium points of irregular-shaped minor celestial bodies. Astrophys. Space Sci. 353(1), 105-121.

Werner, R., 1994. The gravitational potential of a homogeneous polyhedron or don't cut corners. Celest. Mech. Dyn. Astron. 59(3), 253-278.

Werner, R., Scheeres, D. J., 1996. Exterior gravitation of a polyhedron derived and compared with harmonic and mascon gravitation representations of asteroid 4769 Castalia. Celest. Mech. Dyn. Astron. 65(3), 313-344.

Yu, Y., Michel, P., 2018. Ejecta cloud from the aida space project kinetic impact on the secondary of a binary asteroid: II. fates and evolutionary dependencies. Icarus. 312(15), 128-144.

Yu, Y., Richardson, D. C., Michel, P., 2017. Structural analysis of rubble-pile asteroids applied to collisional evolution. Astrodynamics, 1(1), 57-69.